\documentclass[fleqn,usenatbib]{mnras}

\usepackage{newtxtext,newtxmath}

\usepackage[T1]{fontenc}

\DeclareRobustCommand{\VAN}[3]{#2}
\let\VANthebibliography\thebibliography
\def\thebibliography{\DeclareRobustCommand{\VAN}[3]{##3}\VANthebibliography}

\usepackage{graphicx}	
\usepackage{amsmath}	
\usepackage{subfig}

\title[Characteristics of Edge-on Gaseous Counter-rotating Galaxies]{SDSS IV MaNGA: Characteristics of Edge-on Galaxies with 
a Counter-rotating Gaseous Disk}

\author[Minje Beom et al.]{
Minje Beom,$^{1}$\thanks{E-mail: beom@nmsu.edu}
Dmitry Bizyaev,$^{2,3,4}$
Ren\'e A. M. Walterbos$^{1}$
and Yanmei Chen$^{5,6,7}$
\\
$^{1}$Astronomy Department, New Mexico State University, Las Cruces, NM 88003, USA\\
$^{2}$Apache Point Observatory and New Mexico State University, Sunspot, NM 88349, USA\\
$^{3}$Sternberg Astronomical Institute, Moscow State University, Universitetskiy prosp. 13, 119992, Moscow, Russia\\
$^{4}$Special Astrophysical Observatory, Russian Academy of Sciences, 369167 Nizhnij Arkhyz, Russia\\
$^{5}$School of Astronomy and Space Science, Nanjing University, Nanjing 210093, China\\
$^{6}$Key Laboratory of Modern Astronomy and Astrophysics (Nanjing University), Ministry of Education, Nanjing 210093, China\\
$^{7}$Collaborative Innovation Center of Modern Astronomy and Space Exploration, Nanjing 210093, China
}

\date{Accepted 2022 May 23. Received YYY; in original form ZZZ}

\pubyear{2022}

\begin{document}
\label{firstpage}
\pagerange{\pageref{firstpage}--\pageref{lastpage}}
\maketitle

\begin{abstract}
Counter-rotating components in galaxies are one of the most direct
forms of evidence for past gas accretion or merging.
We discovered ten edge-on disk gaseous counter-rotators in 
a sample of 523 edge-on galaxies identified in the final MaNGA 
(Mapping Nearby Galaxies at APO) IFU sample. The counter-rotators tend to located in small groups.
The gaseous counter-rotators have intermediate stellar masses and
and located in the green valley and red sequence of the color magnitude diagram. 
The average vertical extents of the stellar and ionized gas disks are the same as for the rest of the sample
while their radial gas and stellar distributions are more centrally concentrated.
This may point at angular momentum loss 
during the formation process of the counter-rotating disks.
The counter-rotators have low gas and dust content, weak emission line strengths, 
and low star formation rates.
This suggests that the formation of counter-rotators 
may be an efficient way to quench galaxies. 
One counter-rotator, SDSS
J080016.09+292817.1 (Galaxy F), has a post starburst region and a possible
AGN at the center.
Another counter-rotator, SDSS J131234.03+482159.8 (Galaxy H), is identified as a potential on-going galaxy interaction 
with its companion satellite galaxy, a gas rich spiral galaxy. 
This may be representative case of  
a gaseous counter-rotator forming through a merger origin. However, tidal distortions expected in mergers are only found in a few of the galaxies and we cannot rule out direct gas accretion as another formation mechanism.
\end{abstract}




\section{Introduction} \label{sec:intro}
A counter-rotating galaxy is a galaxy containing a component 
with opposite angular momentum to the main stellar disk.
The counter-rotating component can consist of stars or gas,
referred to as a stellar or gaseous counter-rotating galaxy, respectively.
Counter-rotators are interesting galaxies
since the components' opposite angular momenta can not be produced 
by secular evolution. 
Many studies have searched for galaxies with counter-rotating components
\citep{Bertola1985,Galletta1987,BWK,Rix1992,Bertola1992}. 
In these early studies, countor-rotators were generally identified
from long slit spectroscopic observations or 21-cm observations 
(gaseous counter-rotators; \citet{Galletta1987,BWK,Bertola1992}, 
stellar counter-rotators; \citet{Bertola1985,Rubin1992}, 
and see also many further references in the review papers of 
\citet{Galletta1996,BertolaCorsini1999,Corsini2014}).
The early studies showed counter-rotators to be relatively rare galaxies
with a higher rate among early type galaxies 
\citep{Bertola1992,Kuijken1996,KannappanFabricant2001,Pizzella2004}.
IFU (Integral Field Unit) surveys have reported not only counter-rotators, 
but galaxies in which the stellar and gaseous disks 
may be tilted from 30 to 150 degrees. We refer to these as `misalignment galaxie'.
For example, \citet{Davis2011} using $\rm ATLAS^{3D}$
reported early type misalignment galaxies with their spin parameter.
\citet{Bryant2019}, based on the SAMI survey, 
reported several tens of misalignment galaxies 
with a wide range of morphological types.
\citet{Li2021} based on MaNGA (Mapping Nearby Galaxies at APO) 
MPL-8 (MaNGA Product Launches 8th version) identified hundreds of misalignment galaxies 
with morphological features indicating merging or interaction.
In addition, \citet{Xu2022} discuss the global characteristics of 
misalignment galaxies found in MaNGA MPL-10.  

This study is focusing on gaseous counter-rotating galaxies
whose misalignment angle between the rotating stellar and gaseous disks 
is greater than 150 degree. 
Absent the presence of on-going galaxy interaction signatures 
(i.e. morphological changes such as tidal tails) or evidence of
environmental harassment (e.g. tidal tails or stripping), 
counter-rotating components are one of the most direct forms of 
evidence for past gas accretion or merging.
As a result, counter-rotating galaxies are key markers of 
hierarchical galaxy evolution. 
This is because a ‘self gravitating’ evolved system, 
can not make a component with counter-rotating angular momentum. 
In particular, gaseous counter-rotating galaxies may be also 
interesting as a possible evolutionary stage 
in the transition to quiescent galaxies.
The formation process of the counter-rotators results in 
the angular momentum lose in the gas component 
by the hydrodynamic friction \citep{Chen2016,Jin2016}.
It leads an efficient process for driving gas from the disk to the center,
in particular if a pre-existing co-rotating gas disk were present.

This paper is the first study about gaseous counter-rotators 
identified among all targets in MaNGA. 
In this paper, we will focus on edge-on counter-rotators. 
For these special cases, there is no ambiguity in identifying 
co-planar disks and the perspective allows measurement of 
the radial and vertical distributions 
of the gas and stellar components.
We discuss the common characteristics of the counter-rotators 
found in the IFU data, supplemented with information 
from archival data and new observations. 
We relate their properties to possible 
formation processes for gaseous counter-rotators. In particular, the acquisition of the gas may have resulted in a centrally concentrated distribution of gas, central star formation activity, and AGN feeding due to angular momentum loss of the gas.
All these processes make gaseous counter-rotators interesting objects to study
the depletion of the gas component. 
Counter-rotating gas may be accompanied by counter-rotating stars if the accreted gas is dense enough to sustain star formation, or in case of a merger with a gas rich dwarf galaxy. We did not detect evidence for such components in our sample in the MaNGA pipeline products, but placing limits would require more detailed modeling of the spectra. Counter-rotating stellar components in MaNGA galaxies were recently studied by \citet{Becaxqua2022,Bao2022}. The galaxies in our sample do not overlap with objects in their papers.

In section 2, we will present our sample selection, the MaNGA IFU data, archival data and deep imaging using the Apache Point Observatory (APO) 3.5m telescope. 
In section 3, we will present photometric and spectroscopic properties of the counter-rotating galaxies. 
In section 4, we show the connection between the physical properties of counter-rotating galaxies and discuss possible formation mechanisms.
In this paper, a Hubble constant of $70\ \rm km~s^{-1}~Mpc^{-1}$ $(h=0.7)$ is assumed.

\begin{table*}
\centering
\caption{The list of the gaseous counter-rotating galaxies}
\label{tab:list}
\begin{tabular}{ccccccccc}

\hline
MaNGA ID & Plate ID & SDSS ID &
Name used$^{a}$ &
RA & Dec. & z$^{b}$ &
Distance$^{b}$ &
Stellar mass$^{b}$\\
& & & in this paper & 
$\rm [deg.]$ & $\rm [deg.]$ &  & $\rm [Mpc]$  & $\rm[log~ M_{\star}/M_{\odot}]$ \\
\hline
1-225    & 11759-1902 & SDSS J094323.39+004032.6 & A & 145.848 &  0.6757 & 0.02660 & 114 & 9.92 \\
1-382818 & 9491-6102  & SDSS J075943.28+182803.7 & B & 119.930 & 18.4677 & 0.03784 & 162 & 9.94 \\
1-409368 & 12484-1901 & SDSS J130738.44+312913.0 & C & 196.910 & 31.4870 & 0.02615 & 112 & 9.97 \\
1-178790 & 8623-3702  & SDSS J204129.46+000756.1 & D & 310.373 &  0.1323 & 0.02694 & 115 & 10.03 \\
1-261224 & 8334-3703  & SDSS J141255.25+391845.5 & E & 213.230 & 39.3127 & 0.02507 & 107 & 10.10 \\
1-146344 & 8938-6102  & SDSS J080016.09+292817.1 & F & 120.067 & 29.4714 & 0.04527 & 194 & 10.13 \\
1-561039 & 7957-6102  & SDSS J171305.05+351607.0 & G & 258.271 & 35.2686 & 0.02566 & 110 & 10.28 \\
1-576537 & 8465-12704 & SDSS J131234.03+482159.8 & H & 198.142 & 48.3666 & 0.05581 & 239 & 10.65 \\
1-386903 & 12487-3701 & SDSS J091135.65+300439.1 & I & 137.899 & 30.0775 & 0.02598 & 111 & 10.68 \\
1-244629 & 11020-3703 & SDSS J134421.57+555121.3 & J & 206.090 & 55.8560 & 0.03741 & 160 & 10.79 \\
\hline
\multicolumn{9}{l}{$^{a}$The letter designations are taken in order of stellar mass and are used in the paper for brevity.}\\
\multicolumn{9}{l}{$^{b}$The red shift and stellar mass are based on the NSA catalog \citep{NSA}. The values are based on a Hubble constant of $\rm 70~km~s^{-1}~Mpc^{-1}$.}\\
\end{tabular}
\end{table*}

\section{Data and Observations} \label{sec:data}

\subsection{MaNGA data and Sample Selection} \label{sec:sample}

We use the final version of the MaNGA (Mapping Nearby Galaxies at APO) data \citep{mangadr17}.
MaNGA is the survey project in SDSS-IV (Sloan Digital Sky Survey)
using  Integral Field Unit (IFU) observations of galaxies 
to produce spatially resolved spectroscopic data \citep{manga_overview,Blanton2017,Gunn2006,Drory2015,mangadr15,mangadr17}. 
MaNGA observed approximately 10,000 galaxies, 
selected as an unbiased sample in terms of 
stellar mass ($\rm M_{\star} > 10^{9}~M_{\odot}$) and environments
\citep{Law2015,Wake2017,Yan2016a}. 
The MaNGA targets were selected as two main subgroups and two minor subgroups. 
Among the main subgroups, the primary sample of about 5,000 galaxies was selected 
to observe the central part out to $\rm 1.5~R_{eff}$. 
The secondary sample of the main subgroups includes about 3,300 galaxies, 
observed by the IFU bundle out to $\rm 2.5~R_{eff}$. These samples were selected without bias and covers a wide range in galaxy masses.
The first of the two minor subgroups is a "color-enhanced sample" of about 1,700 galaxies.
It is selected to include blue massive galaxies and green valley galaxies 
which are important to study the quenching process.
These targets are observed with an IFU coverage of $\rm 1.5~R_{eff}$.
A second minor sample includes about 1000 ancillary targets, selected for various observations using the unique capability of the MaNGA instrument \citep{Yan2016a}.
The IFU bundles of the MaNGA consist of 19 to 127 fibers with hexagonal shape 
covering $12$ to $\rm 32~arcsec$ in diameter on sky. 
The spatial resolution of MaNGA is $\rm 2.5~arcsec$, 
which corresponds to $1.3$ to $\rm 4.5~kpc$ for the primary sample and 
$2.2$ to $\rm 5.1~kpc$ for the secondary sample. 
The BOSS spectrographs used by MaNGA provide spectra from 3600 to 10300~\AA~
with a spectral resolution of about 2100 at 6000~\AA~
\citep{BOSS,Yan2016b}. 
The observations were conducted to reach a S/N in the stellar continuum at $\rm 1~R_{eff}$ of $\rm 14$ to $\rm 35$ per spatial sample, 
which required about 3 hours net integration for each target.


Our sample consists of edge-on disk galaxies selected 
from the MaNGA sample through visual inspection 
using the SDSS (Sloan Digital Sky Survey) images and the ionized gas distributions shown in MaNGA.
We identified 523 edge-on disk galaxies out of 10,590 MaNGA targets 
from the final version of MaNGA Product Launches 11 (MPL-11) 
as a part of SDSS Data Release 17 (DR17) \citep{mangadr17}.
The edge-on sample was selected based on two considerations: edge-on inclination and 
no clear evidence for ongoing major merging interaction. The assessment was 
based on visual inspection of SDSS images.
This sample selection is similar to that of \citet{BWY17}.
Some  galaxies  have  well  identified  dust  lanes projected 
on  the  galactic  nucleus  confirming  an  edge-on orientation 
but this is not the case for all of them.
Typically, their inclination is estimated to be very close to 90 degrees. 
We rejected galaxies showing obvious signatures of 
ongoing major merging interaction on SDSS optical broad band images, 
such as morphological distortions in the optical image.
We also excluded a few galaxies showing extreme distortion in H$\alpha$
distribution. The goal of the selection was to identify a large sample of normal appearing edge-on galaxies to study their ionized gas distributions and kinematics in radial and vertical directions.
While we excluded galaxies with obvious features suggesting interaction
some galaxies in our sample may be interacting, e.g.
with a nearby dwarf companion while not showing distortion in the stellar light distribution.
We did not take into account IFU coverage in the selection, hence the radial coverage is not the same for all objects but in all cases the coverage is sufficient to reach well into the disk areas.
Based on the NSA (NASA Sloan Atlas) photometric properties of \citet{NSA}, 
the stellar mass and the absolute magnitude of the sample are
from $3\times10^{8}$ to $2\times10^{11}$ and $-15$ to $-21$ in r band.
The redshift range is from $0.01$ to $0.06$, 
which corresponds to about $50$ to $\rm 250~Mpc$
(half of sample in the range $0.02$ to $0.03$).

We also use MaNGA VACs (value added catalogs), in particular the HI-MaNGA and GEMA (Galaxy Environment for MaNGA) results.
The HI-MaNGA survey is an on-going project to ascertain HI properties 
for the MaNGA targets using new observations by the GBT (Green Bank Telescope) 
and archival observation data from ALFALFA (The Arecibo Legacy Fast ALFA Survey) \citep{Haynes2018,Masters2019,Goddy2020,Stark2021}.
The GEMA is a catalog analyzing the 
environments of the MaNGA targets using the method of \citet{Argudo2015}.

\subsection{MaNGA IFU Data Processing and Uncertainty} \label{sec:manga}

We use the data products from the MaNGA DRP 
(Data Reduction Pipeline, version 3.1.1) 
and DAP (Data Analysis Pipeline, version 3.1.0) \citep{DRP,DAP1,DAP2,DAP3}.
All figures and results related to the spatial distributions of stellar and gas components
are based on data with the default spaxel binning 
(spaxels are 0.5").
For the stellar component, the surface brightness distributions in the in g-band
are shown only for spaxels where the signal-to-noise ratio (SNR) is above 3.
For the gas component, 
we show the surface brightness distributions of the H$_\alpha$ with a SNR cut of 3
estimated by the `non-parametric flux sum' method 
of the MaNGA DAP, which is called `SFLUX' in the DAP data.

Based on the dependence of the kinematic uncertainty on SNR, 
we apply SNR cuts of 3 for the velocity fields, 
and 6 for the velocity dispersion fields.
The velocity uncertainty tends to be higher in the stellar component than in the gas component,
while the uncertainty of the velocity dispersion tends to be higher in the gas component. 
We also note that the kinematic features discussed in the paper 
are present in the MaNGA DAP output made with different stellar libraries in the spectrum fitting\footnote{The MILES and MaSTAR stellar libraries were used in the MaNGA DAP data of SDSS DR15 and SDSS DR17, respectively. Details may be found in \citet{DAP1,DAP3}}.
We take the kinematic values of the gas component
estimated from the`gaussian integrated flux' method of the MaNGA DAP. 
This method fits major emission lines together 
in order to minimize uncertainties \citet{DAP1,DAP2}.
It produces emission line properties in the `MAP' file 
with names of `GVEL', `GSIGMA' and `GFLUX'.

We also estimate emission line strengths from integrated spectra 
of the whole galaxy and of the central $\rm 2.5~arcsec$ region using pPXF 
(Penalized Pixel-Fitting, \citet{pPXF}). 
The integrated spectrum of the whole galaxy is summed from all valid spaxels 
where the mean surface brightness in g band or H$\alpha$ wavelength 
has signal-to-noise ratio higher than $3$.
The central spectrum is integrated within the central $\rm 2.5~arcsec$ region 
and is used to look for AGN signatures in emission lines. Of course, for edge-on galaxies an AGN signature may not be visible depending on the extinction and this spectral test is not conclusive, although most of our objects tend to have low dust content as we will show.
The results of these integrated spectra are used for the diagnostic diagrams and the star formation rate figure
in the section \ref{sec:ionized_gas} and \ref{sec:SFR}.
For figures which are referred in the other sections, 
we use the DAP products to show the distribution of 
stellar and gas properties.

\begin{figure*}
\centering

\includegraphics[scale=0.36]{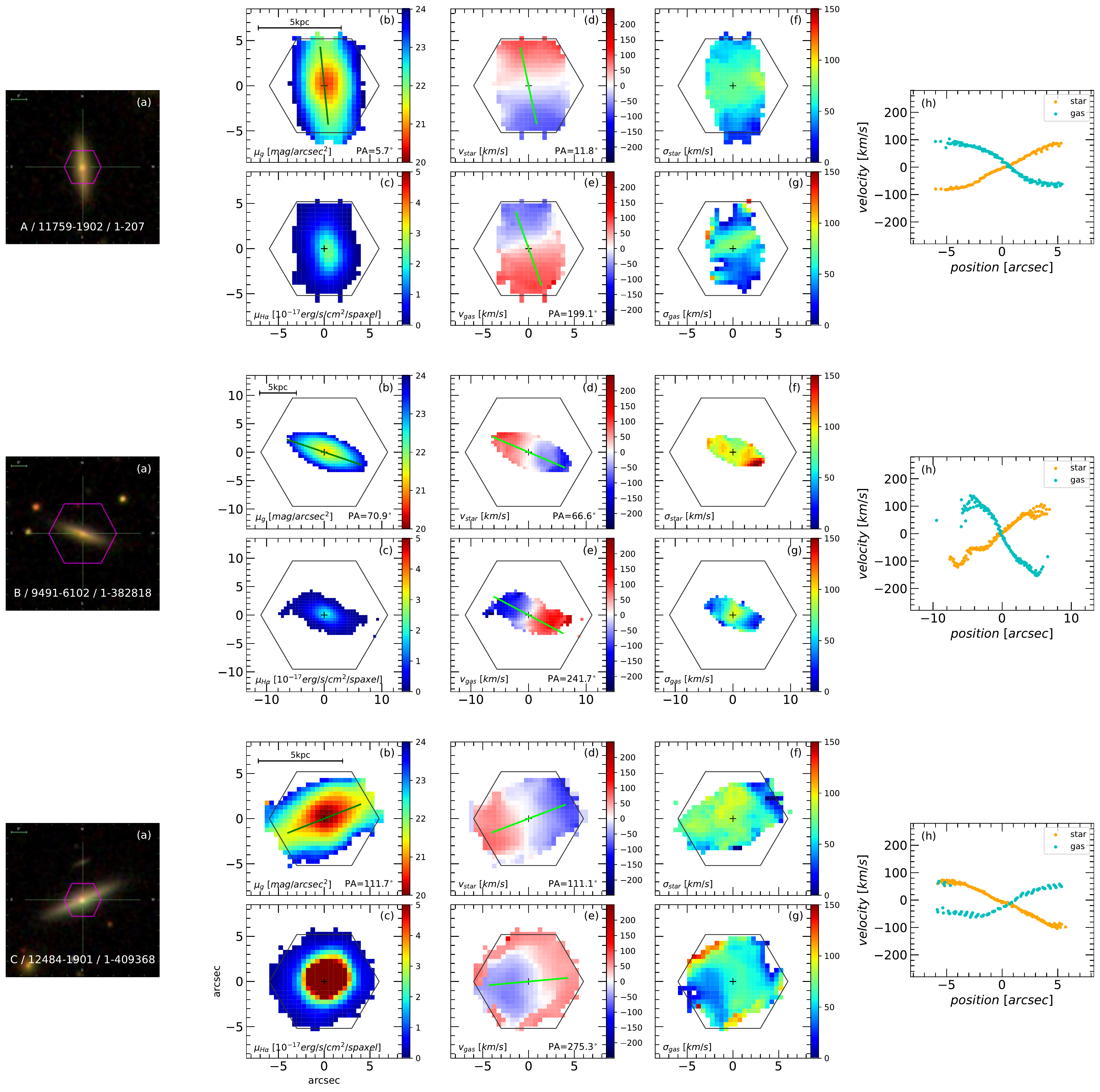}
\caption{Counter-rotating galaxies A, B, and C. Each counter-rotating galaxy is shown with the SDSS broadband image (a), surface brightness (b and c), kinematics (d to g), and  position-velocity diagram (h). 
(a): SDSS broadband color image with each IFU coverage as a purple hexagon.
(b): the surface brightness of the stellar component in units of the mean magnitude in g-band per spaxel. 
The green line in the (b) panel indicates the photometric semi-major axis.
(c): the surface brightness of the ionized gas in H$\alpha$ emission line flux per spaxel. 
(d) and (e): the line-of-sight velocity in the stellar absorption lines and the ionized gas emission lines, respectively.
The light green line in the (d) and (e) panels indicates the kinematic major axis.
(f) and (g): the velocity dispersion in the stellar absorption lines and the ionized gas emission lines, respectively.
The black hexagon in the (b) to (g) panels is the IFU coverage.
(h): The position-velocity diagram; velocities along the kinematic major axis of the stellar (orange dots) and gaseous (light blue dots) disks.
\label{fig:kinematics_fig1a}}
\end{figure*}

\begin{figure*}\ContinuedFloat
\centering
\includegraphics[scale=0.36]{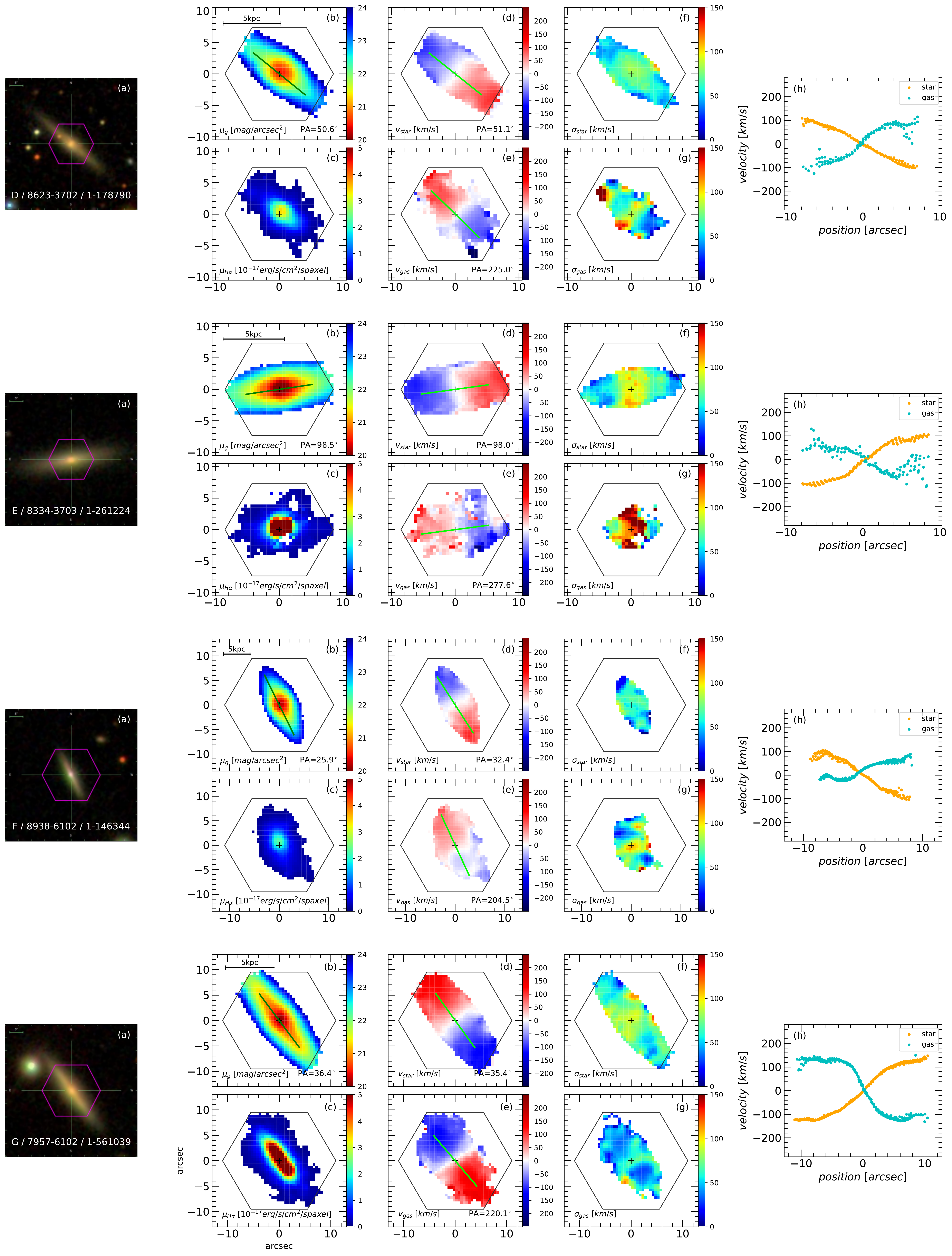}
\caption{(Continued) Counter-rotating galaxies D, E, F and G. 
\label{fig:kinematics_fig1b}}
\end{figure*}

\begin{figure*}\ContinuedFloat
\centering
\includegraphics[scale=0.36]{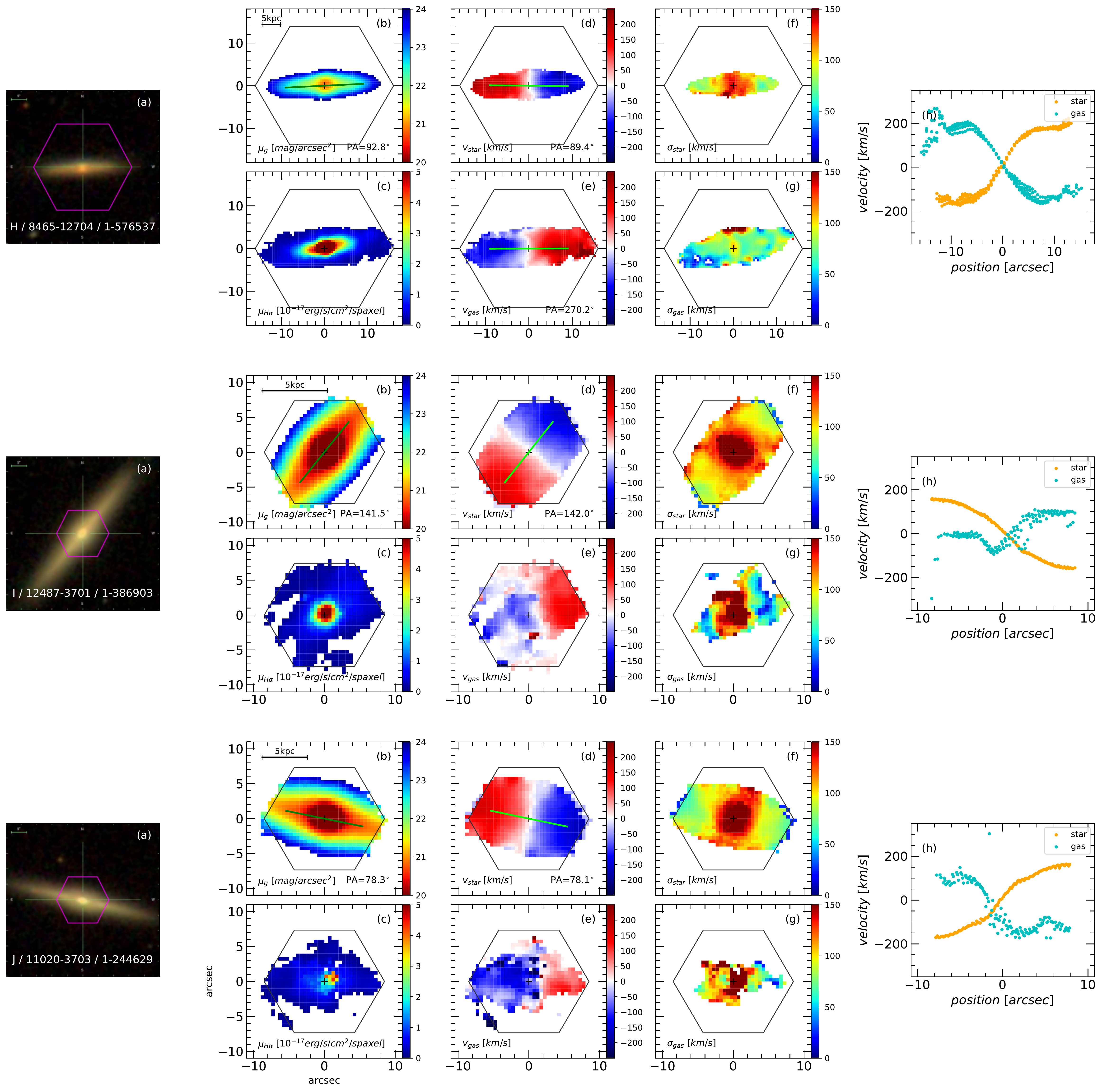}
\caption{(Continued) Counter-rotating galaxies H, I, and J. See Figure~\ref{fig:kinematics_fig1a} for caption.
The kinematic major axes in the gas disks of Galaxy I and J are not plotted
because they are not properly constrained due to weak emission strength 
and an irregular shape of their gas disks. 
\label{fig:kinematics_fig1c}}
\end{figure*}

\subsection{Archival Data} \label{catalogs}

This study uses archival data for photometric data and physical properties of galaxies.
We obtain the photometric properties of the edge-on galaxies 
from the NSA catalog (NASA Sloan Atlas) and 
the WISE (Wide-field Infrared Survey Explorer) all-sky database \citep{NSA,WISE}.
Since MaNGA targets were selected based on NSA catalog information,
all edge-on disk galaxies are matched in the NSA catalog.
However, some edge-on galaxies were not detected with WISE.
We restricted analyses to the 472 galaxies with available W3 magnitude ($\rm \lambda =12 \micron$) for WISE color-color diagram (Fig.~\ref{fig:Dust}) 
and star formation rate using W3 band (the right panel of Fig.~\ref{fig:SFR}).   
Of these 472 galaxies, 276 have observed W4 magnitude ($\rm \lambda = 22 \micron$) and in this case we estimated the
star formation rates using a combination of H$\alpha$ and $W4$ luminosities.
We investigate HI gas contents from HI-MaNGA \citet{Masters2019,Stark2021} and
bulge and disk decomposition information from \citet{Simard2011}.
In addition, we check the galaxy morphology and search for nearby companions 
using DECaLS and PanSTARR images which are deeper or sharper than SDSS broad band images \citep{PanSTARR, DeCALS}.

\subsection{APO 3.5m Deep Imaging Observations} 
\label{sec:apo35}
All the counter-rotating galaxies were observed 
with the ARCTIC imager on the 3.5m ARC telescope at the
Apache Point Observatory in December 2017, March 2018, January 2020 and January 2021. 
The objects were observed under photometric
conditions with a median seeing of $\rm 1.1~arcsec$.
Each galaxy was imaged with three frames of 15 or 20 minutes exposures
in the SDSS r-band, with spatial dithering between them.
Calibration frames (biases, darks, and sky flats)
were obtained every observing night.
The images were reduced and co-added using standard IRAF packages.
We performed photometric calibration via stars in the wide
fields with known magnitudes in the SDSS passbands. Typically,
we selected ten stars in the field that were faint enough for their counts to fall in the linear regime of the CCD response.
The deep images allow us to search for extended features around the 
galaxies down to about $\rm 28.3~mag~arcsec^{-2}$ in the r-band.

\section{Results} \label{sec:result}
\subsection{Identification of counter-rotating Galaxies} \label{sec:crgal}

We found ten counter-rotating disk galaxies in our edge-on disk galaxy sample 
through visual inspection of the MaNGA DAP output. 
In the visual inspection, we defined a counter-rotator if it had a misalignment 
higher than $150~degrees$ in the position angle of the stellar and gas disks. 
All counter-rotating edge-on disk galaxies have their gas and stellar disks co-planar or very close to co-planar (from $170$ to $\rm 180~degree$).
Fig~\ref{fig:kinematics_fig1a} 
shows for all objects a color image from the SDSS survey, 
the distribution of the stellar and ionized gas, and 
their velocity and velocity dispersion fields.
The counter-rotating galaxies are listed in Table \ref{tab:list}. 
To simplify naming them in the rest of the paper, we use letter designations from A to J, 
in order of the galaxies' stellar masses.

\begin{table*}
\centering
\caption{Physical properties and Environment}
\label{tab:photometric}
\begin{tabular}{ccccccccccc}

\hline
Name & $\rm M_{*}$ $^{a}$ & $\rm R_{eff,p}$ $^{a,b}$ &
$\rm M_{r}$ $^{a,c}$ & $\rm g-r$ $^{a,c}$ & $\rm B/T_{r}$ $^{d}$ & $\rm n_{b}$ $^{d}$ & 
SFR$^{e}$ & Metallicity$^{g}$ & 
$\rm N_{gal.}$ $^{h}$ & Closest galaxy\\
 & $\rm [10^{10}$ $\rm M_{\odot}]$ & $\rm [kpc]$ & 
$\rm [mag.]$ & $\rm [mag.]$ &  & &
$\rm [M_{\odot}/yr]$& $\rm 12+log(O/H)$& 
& $\rm [D_{proj.},\Delta v_{r}]$ $^{j}$ \\
\hline
A & 0.829 & 1.60 & -18.68 & 0.81  & 0.32 & 5.64 &0.170 &     8.64&     2 & $\rm 529~kpc, -405~km~s^{-1}$ \\
B & 0.878 & 1.95 & -17.95 & 0.82  & 0.16 & 4.09 &0.393 &     8.64&     4 & $\rm 70.6~kpc, 344~km~s^{-1}$ \\
C & 0.931 & 1.50 & -18.87 & 0.62  & 0.44 & 4.42 &0.277 &     8.64&     1 & - \\
D & 1.069 & 1.60 & -18.10 & 0.65  & 0.44 & 1.38 &0.392 &     8.68&     1 & - \\
E & 1.270 & 1.82 & -18.37 & 0.72  & 0.52 & 3.69 &0.378 &     8.72&     5 & $\rm 134~kpc, 380~km~s^{-1}$ \\
F & 1.357 & 2.01 & -18.87 & 0.59  & 0.55 & 5.59 &$\leq$0.621$^{f}$ &     8.51& 4 & $\rm 129~kpc, 249~km~s^{-1}$ \\
G & 1.904 & 2.21 & -18.73 & 0.77  & 0.20 & 4.71 &0.245 &     8.72&     4$^{i}$ & $\rm 351~kpc, -50.4~km~s^{-1}$ \\
H & 4.433 & 4.65 & -19.71 & 0.82  & 0.35 & 4.59 &0.980 &     8.58&     2 & $\rm 95.0~kpc, -82.7~km~s^{-1}$ \\
I & 4.734 & 1.78 & -20.50 & 0.73  & 0.49 & 4.50 &0.486 &     8.67&     5 & $\rm 111~kpc, 267~km~s^{-1}$   \\
J & 6.206 & 2.49 & -20.11 & 0.72  & 0.47 & 5.79 &0.549 &     -&     4 & $\rm 255~kpc, -21.1~km~s^{-1}$   \\
\hline
\multicolumn{11}{l}{$^{a}$Properties are taken from the NSA catalog \citep{NSA} }\\
\multicolumn{11}{l}{$^{b}$The effective radius is the radius containing 50\% light of the Petrosian flux.}\\
\multicolumn{11}{l}{$^{c}$The uncertainty of the g and r magnitudes are about 0.020 to 0.025.}\\
\multicolumn{11}{l}{$^{d}$The bulge to total light ratio in the r band and the sersic index of the bulge are taken from \citet{Simard2011}.}\\
\multicolumn{11}{l}{$^{e}$The star formation rate and the specific star formation rate are estimated from the stellar mass and 
the star formation rate traced by H$\alpha$ luminosity }\\
\multicolumn{11}{l}{ with WISE W4-band luminosity for Galaxy C, E, and G, and W3-band luminosity for the others. }\\
\multicolumn{11}{l}{$^{f}$The values with `$\leq$' symbol indicate upper limits based on upper limits in the WISE.}\\
\multicolumn{11}{l}{$^{g}$The gas phase metallicity is estimated with O3N2 indicator 
and normalized with the solar metallicity of 8.69. Details are in the section \ref{sec:gas_metallicity}. }\\
\multicolumn{11}{l}{Galaxy J is not estimated due to its weak H$\beta$ emission line strength.} \\
\multicolumn{11}{l}{$^{h}$We counted the group member galaxies having a spectroscopic redshift in SDSS within a $\rm 600~kpc$ radius field and $\rm \pm 500~km~s^{-1}$ in radial velocity. }\\
\multicolumn{11}{l}{$^{i}$There are several nearby galaxies which are not counted in the number due to not having a spectroscopic redshift. Details are in the section \ref{sec:environment}. }\\
\multicolumn{11}{l}{$^{j}$The projected distance at the redshift of the targets and the radial velocity difference. }\\
\end{tabular}
\end{table*}

\begin{figure*}
\centering
\includegraphics[scale=0.52]{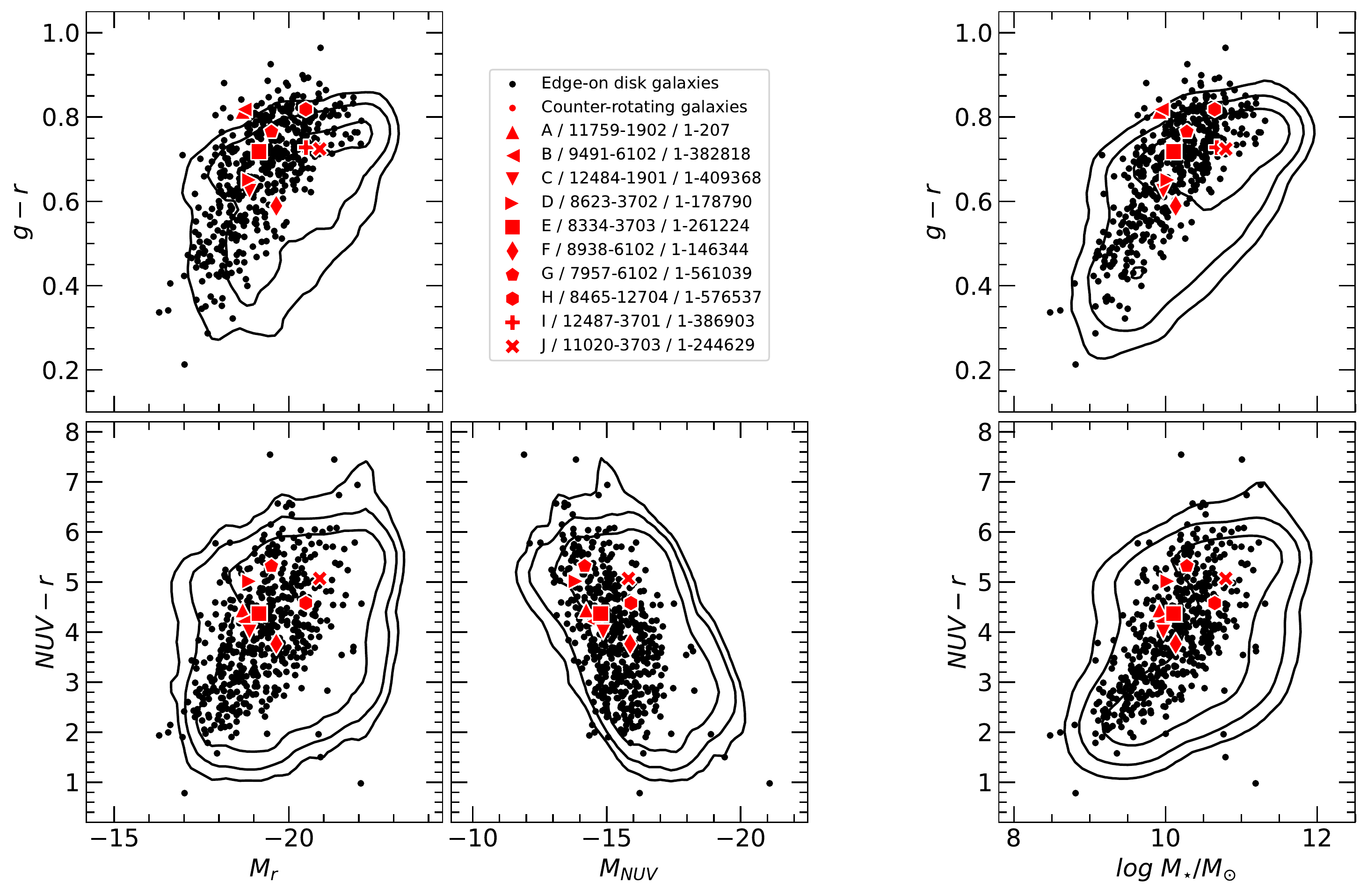}
\caption{The color-magnitude diagram and the color-stellar mass diagram showing all edge-on galaxies (black) and the counter-rotators (red).
The contours show the distribution for all morphological types and inclinations in the MaNGA sample. 
The left three panels show visible and NUV colors. 
The right two panels show color versus stellar mass. 
Galaxy H is not plotted on the three bottom panels 
due to a high uncertainty in NUV magnitude.
The counter-rotating galaxies are located in the red sequence or green valley regime 
of the color-magnitude diagrams.
 \label{fig:CMD}}
\end{figure*}

\begin{figure}
\centering
\includegraphics[scale=0.40]{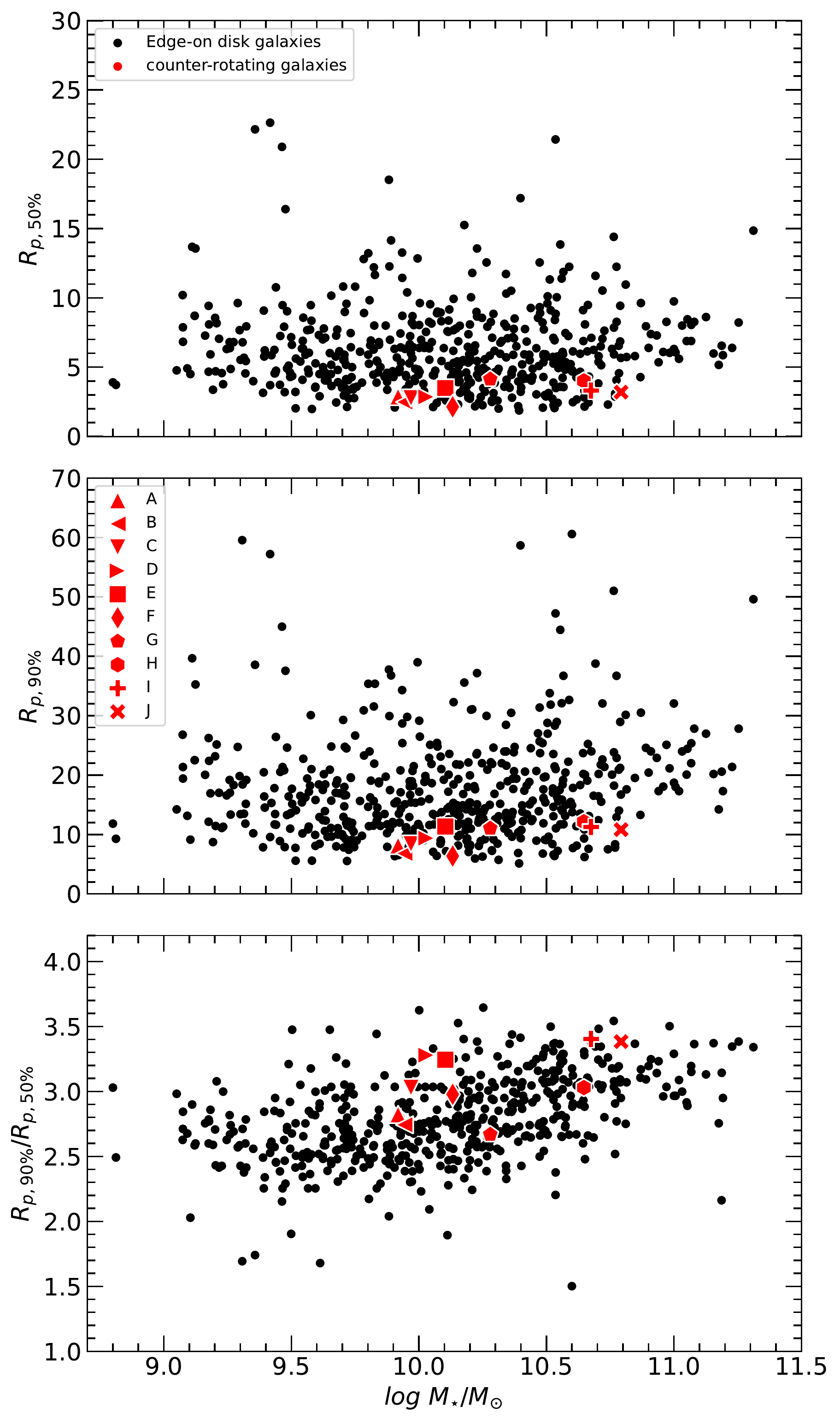}
\caption{The 50\% light (effective) and 90\% light radii 
and concentration index $\rm (R_{p,90}/R_{p,50})$ in the r band
plotted against the stellar masses of the edge-on galaxies.
The petrosian 50\% and 90\% light radii are taken from the NSA catalog \citep{NSA}. 
The counter-rotating galaxies appear to have small 50\% and 90\% light radii. \label{fig:radius}}
\end{figure}

\subsection{Optical Photometric properties} \label{sec:photometric}

The photometric properties of the counter-rotating galaxies 
are summarized in Table~\ref{tab:photometric}. 
The major photometric parameters in the table are taken from the NSA catalog \citep{NSA}, 
while the bulge-to-disk light ratios and the sersic index come from \citet{Simard2011}. 
The counter-rotating galaxies have absolute magnitudes of 
$-18$ to $-20$ in the r-band.
These correspond to stellar masses of 
$8.3\times10^{9}$ to $\rm 6.2\times10^{10}$ $M_{\odot}$.
This places the objects in an intermediate stellar mass range,
which is smaller than the characteristic mass in the galaxy mass function 
for nearby galaxies $\rm 2\times10^{11}~M_{\odot}$ \citep{Baldry2008}.

The counter-rotating galaxies appear not to be blue spiral galaxies. 
Fig.~\ref{fig:CMD} shows CMDs and color versus stellar mass plots for MaNGA galaxies. 
As seen in all panels in the figure, the distribution of the 
edge-on galaxies (black dots) are shifted to a redder color regime
in comparison to the distribution of the overall MaNGA galaxies 
(black contours).
This is likely due to increased extinction in the edge-on perspective.
The inclination effect is more significant for dust rich edge-on disk galaxies 
than for intrinsically less dusty red galaxies.
The bottom edge of the distribution for edge-on galaxies 
(black and red symbols) 
is shifted by about 0.4 dex in g-r CMD,
while the upper edge of the distribution is shifted by about 0.2 dex 
compared to the distribution for all MaNGA galaxies.
With these points in mind, 
the seven counter-rotating galaxies 
might be identified as red sequence or green valley galaxies. 
Even considering the uncertainty due to extinction,
we can conservatively claim that 
the counter-rotating galaxies are likely not blue cloud galaxies 
that are more heavily reddened. 
In agreement with this, the counter-rotators show no evidence 
for strong dust lanes projected against their bulges 
at the resolution limit (one to few kpc) of the SDSS optical images 
(see the first column panels of Fig.~\ref{fig:kinematics_fig1a}.
We also confirmed this with images of PanSTARRS 
(Panoramic Survey Telescope and Rapid Response System) 
and the DESI survey (DECaLS, Dark Energy Camera Legacy Survey) \citep{PanSTARR, DeCALS}. 
In addition, this is in agreement with the low dust content shown 
on the WISE infrared color-color diagram, 
which will be discussed in the next section.

The concentration of the stellar component from 
the radii including 50\% and 90\% light suggests the counter-rotating galaxies are compact.
Fig.~\ref{fig:radius} shows the half light radii (effective radii), 
and the 90\% light radii in the r band plotted against the stellar masses.
In the NSA catalog the radii were calculated 
from a petrosian fitting model \citep{NSA}. 
The figure also shows the concentration index, defined as
the ratio of the radii containing 90\% and 50\% fluxes. 
As seen in Fig.~\ref{fig:radius}, all counter-rotating galaxies 
are appeared as small galaxies in the both of effective and 90\% light radii
compared to those edge-on disk galaxies of similar stellar masses.
Although Galaxies D, E, I and J have high concentration indexes,
their concentration indexes are not small in the ratio of concentration index ($\rm R_{p,90\%}/R_{p,50\%}$), 
which is in a typical range for spiral galaxies.
Nevertheless, all counter-rotating galaxies
have small overall size as well as small effective radii.

Given the edge-on nature, 
it is not easy to classify disk galaxies into S0 and spiral types,
since the spiral arms, one of the main parameter for classification of disk galaxies, are not visible.
\citet{Dominguez_Sanchez2022} derived morphological classifications
for the MaNGA galaxies using a combined visual and machine learning approach. They noted that
the morphological classification for the edge-on galaxies
is uncertain.
We did a careful visual inspection for the counter-rotators. 
The absence of strong dust lanes and of possible enhancements that might be present in an edge-on view along spiral arm segments suggests the counter-rotators could be S0 types rather than spiral galaxies. However, this is uncertain given the limited spatial resolution (0.5 kpc or worse).
An S0 classification may be supported by a lack of distinct large star formation regions, but this can be true for early type spirals too.
We did identify many gaseous counter-rotators 
with morphological features such as a bar, a ring, or spiral arms 
in an intermediate inclination sample \citep{Beom2022}. This suggests it may be unlikely that all the edge-on counter-rotators happen to be S0s.
We also analyzed the bulge-to-total (B/T) light ratio and the concentration index (the ratio of the radii containing 90\% and 50\% of the light; R$_{90\%}$/R$_{50\%}$)
to try to estimate their sub-class. 
Based on the B/T ratio reported on \citet{GrahamWorley2008, KormendyBender2012},
Galaxy B and G would be S(0)c and 
the other eight counter-rotators are S(0)a and S(0)b, if indeed they are all S0s. All the counter-rotators have
a concentration index value larger than 2.7,  
in the range of S(0)a and S(0)b \citep{Kautsch2006}.
Since the B/T ratio and the concentration index are derived 
without considering extinction and inclination effects 
this classification too remains somewhat uncertain.

\begin{figure*}
\centering
\includegraphics[scale=0.6]{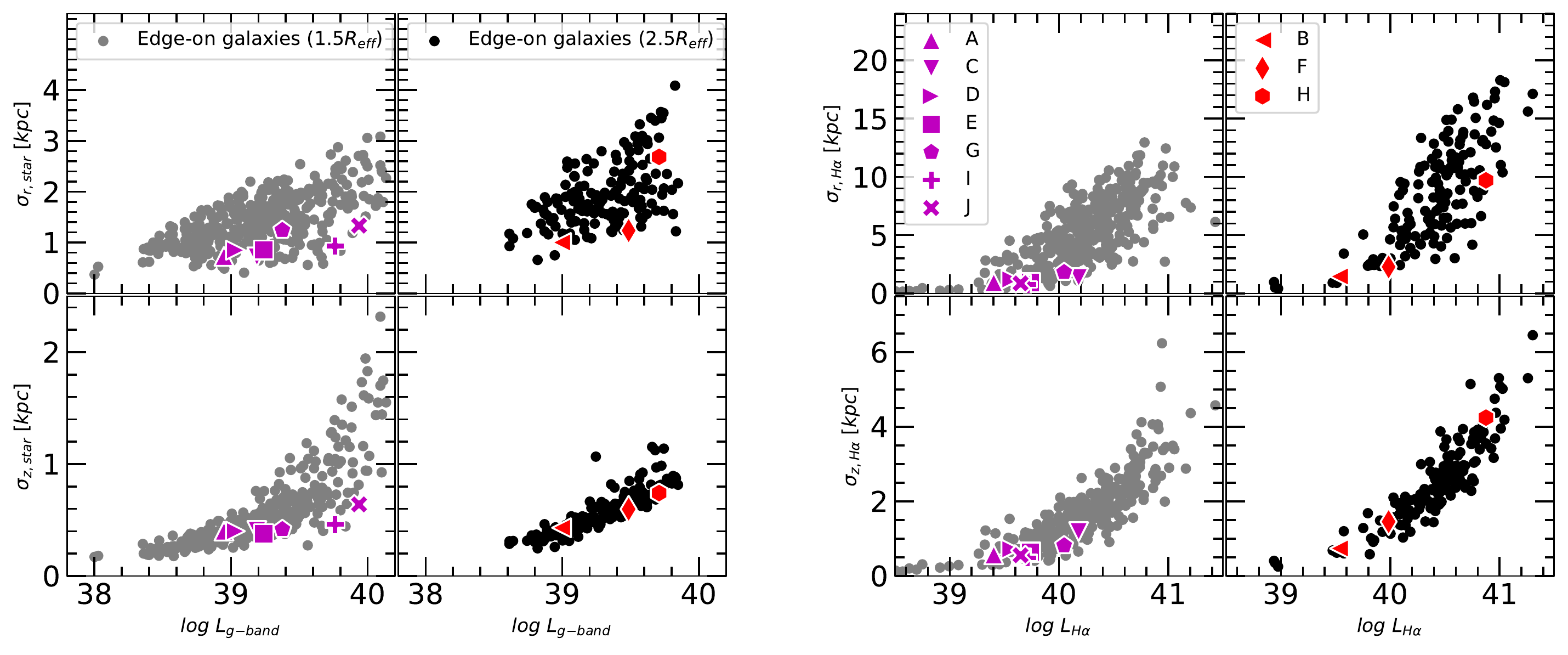}
\caption{The stellar and gaseous extents derived from the 
intensity weighted light distributions in the radial direction and
the vertical direction. 
The left four panels are for the stellar distribution, and 
the right four panels are for the ionized gaseous distribution traced by H$\alpha$.
The upper row panels show the values in the radial direction,
and the bottom row panels show the values in the vertical direction.
The x-axis of each panel is the g-band luminosity
or the luminosity of H$\alpha$ emission line.
Each edge-on galaxy is presented 
as grey or purple symbols for the primary sample of MaNGA ($\rm 1.5~R_{eff}$)
and black or red symbols for the secondary sample ($\rm 2.5~R_{eff}$).
Except for Galaxy H, the counter-rotators show 
a centrally concentrated distribution in both stars and gas, 
especially in the radial direction. 
 \label{fig:X2Y2}}
\end{figure*}

\subsection{The Stellar and the Ionized Gas Distributions} \label{sec:distribution}

The H$\alpha$ panels of Fig.~\ref{fig:kinematics_fig1a}
show the ionized gas distribution 
in the counter-rotating galaxies. The ionized gas emission is
strongest at their center and rapidly decreases with radius.
None of the counter-rotators
show distinct bright HII region complexes in their disks, such as may be seen in normal later type spirals.
Galaxies G and H have strong emission lines out to the middle of the stellar disk.
This extended distribution may still be 
continuously connected to the central emission, 
not a separate region from the center.

We quantitatively evaluated the concentration of 
the ionized gas and the stellar distribution in the vertical and radial directions
in comparison with the other edge-on galaxies.
We calculate the second momentum of the light distibution
using the flux per spaxel as weights in the average. 
This can be expressed as
\begin{equation} 
  \centering
  \langle x^{2} \rangle =\frac{\sum_{i} F_{i} x_{i}^{2}}{\sum_{i} F_{i}} 
  \label{flux-weighted average}
\end{equation}
where x (y) is the coordinate centered on the galaxy nucleus
and $F_{i}$ is the flux contained inside each spaxel (or a spaxel bin). 
The quantity in equation \ref{flux-weighted average} is a measure of the surface brightness extent in the x-axis direction.
This definition is taken from \citet{Emsellem2007,Cappellari2007}.
We rotated each galaxy to align the major axis with the x-axis prior to calculating the moments.
We then obtain
\begin{equation} 
  \centering
  \begin{split}
    \sigma_{r}=\left( \langle x^{2} \rangle \right) ^{1/2}=
    \left(\frac{\sum_{i} F_{i,rot} x_{i,rot}^{2}}{\sum_{i} F_{i,rot}}\right)^{1/2} \\
    \sigma_{z}=\left( \langle y^{2} \rangle \right) ^{1/2}=
    \left(\frac{\sum_{i} F_{i,rot} y_{i,rot}^{2}}{\sum_{i} F_{i,rot}}\right)^{1/2} \\
  \end{split}
  \label{kinematic_major}
\end{equation}
where $F_{i,rot}$ is the H$\alpha$ flux per spaxel.
Similarly, The mean intensity per spaxel in the g-band is used for the stellar component.
$\rm x_{i,rot}$ and $\rm y_{i,rot}$ are the coordinates 
with respect to the galaxy's center after the coordinate rotation.
The subscripts `r' or `z' are used 
to refer to the radial and vertical directions. 
Fig.~\ref{fig:X2Y2} shows the results.
Due to the IFU coverage, the evaluation of the distribution 
in the radial direction is likely underestimated 
for the primary sample of the MaNGA targets. 
Thus, Galaxy B, F, and H (red polygons) should be compared 
within the secondary sample (black dots), while
the other counter-rotators (purple polygons) should be compared 
within the primary sample (grey dots).

As seen the left upper two panels, 
the nine counter-rotators, except for Galaxy H, have 
a small extent in the radial direction for the stellar component, in agreement with  
the small effective radius and the 90\% light radius 
for counter-rotators shown in Fig.~\ref{fig:radius}.
In the bottom left two panels, 
note 
the relation between the vertical extent or the starlight and 
the luminosity (also a proxy for the stellar mass) of the counter-rotators.
The values for counter-rotators are in the lower range for edge-on galaxies.
Galaxy I and J, the most massive two counter-rotators, 
have a noticeably small extent in the vertical distribution.

In the upper right two panels we see that 
the counter-rotators, except for Galaxy H, also
have a small ionized gas extent in the radial direction.
On the other hand, the gas distribution in the vertical direction  
follows the trend that bright galaxy in H$\alpha$ tend to have larger vertical extents and the counter-rotators are not particularly smaller in vertical gas extent.

Galaxy H is the only one counter-rotator showing 
relatively `normal radial extents in both the stellar and the gas components.
We will revisit this in the discussion section.

\subsection{Dust and HI Contents} \label{sec:Dust}

\begin{figure}
\centering
\includegraphics[scale=0.35]{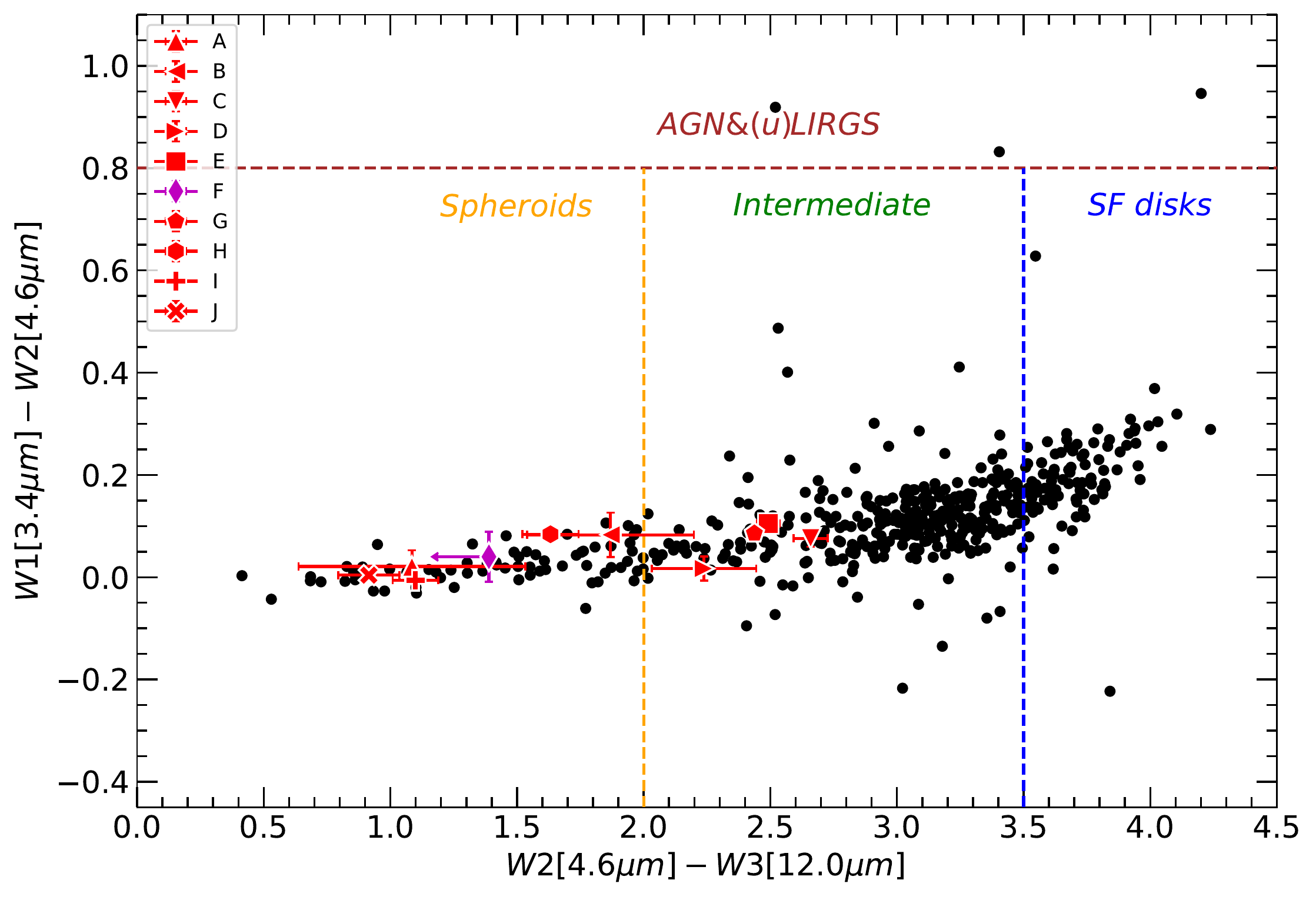}
\caption{The color-color diagram using W1-W2 and W2-W3 colors
from WISE infrared observations. 
The black dots are edge-on galaxies detected 
in the three bands. 
The counter-rotating galaxies are shown as red or purple polygons. 
The lines indicating classes are taken from \citet{Jarrett2013}. 
Galaxy F has an upper limit value of W3 brightness so that 
its W2-W3 color for the x-axis is is upper limit, 
which is present as purple. 
 \label{fig:Dust}}
\end{figure}

WISE (WIDE-field Infrared Survey Explorer) data may be used to access 
the dust content of the galaxies. 
WISE took data in four infrared bands W1, W2, W3, and W4 
(3.4, 4.6, 12 and 22~$\micron$)
with a spatial resolution of 6.1, 6.4, 6.5, and 12~arcsec \citep{WISE}.
\citet{Jarrett2011,Jarrett2017} showed the color-color diagram using 
W1, W2 and W3 bands can be used to assess dust content in the galaxies.
Fig~\ref{fig:Dust} shows the W1-W2 vs. W2-W3 color-color diagram
for the edge-on disk galaxies. 
Here the x-axis (W2-W3) is a proxy for dust contents. 
The counter-rotating galaxies are located in the
`Spheroids' and left part of `intermediate' regimes on the diagram, 
consistent with low dust content. 
Their low dust content can also be inferred from the 
low luminosities in the W3 and W4 bands, 
discussed further in the context of star formation properties 
in section \ref{sec:SFR}. 
In addition, \citet{Jarrett2011,Jarrett2017} also show that 
strong AGN galaxies would be loacted in the upper regime of the diagram. 
None of the counter-rotators appear as strong AGN in the diagram.

Four counter-rotators, Galaxy B, D, F and G,
have been observed so far by HI-MaNGA survey
\citep{Haynes2018,Masters2019,Goddy2020,Stark2021}. 
Galaxy G is the only one detected among the four galaxies.
The HI mass of Galaxy G is $3.86\times10^{9}$ M$_{\odot}$, which 
corresponds to 0.2 in M$_{HI}$/M$_{\star}$.
The non-detections for the other three targets (Galaxy B, D, and F)
seem to indicate small HI masses.
Owing to the large beam size of 9 arcmin of GBT, moreover,
the HI mass of the HI-MaNGA might include also HI of nearby galaxies;
Galaxy G has candidate nearby galaxies in the optical images,
we refer to the section \ref{sec:environment}).
Galaxies B and F each have a companion galaxy
within the GBT beam, located at a projected distance of 
about 1.5 and 2~arcmin, respectively.
With these considerations the HI-MaNGA observation data
for Galaxy B, D, F and G are in agreement with low gas content.

\subsection{Kinematics} 
\label{sec:Kinematics}

We determined the kinematic major axis and 
the rotational velocity along the major axis and searched for any disturbances in the velocity and velocity dispersion.
Due to the edge-on perspective, we can consider the line of sight velocity 
a reasonable approximation to the rotational velocity. 
However, line-of-sight projection effects through a rotating disk imply that the fitted radial velocity is not exactly the tangential velocity at each radius.
We focus on the comparison of the kinematics between stellar 
and gas components rather than the absolute values of the rotational velocity.

\subsubsection{Kinematic Major and Minor Axes}
\label{sec:kinematic_axes}

We derive the kinematic major axis by fitting the optimum kinematic position angle (PA) 
at which the observed velocity field matches a model velocity field. 
The free parameters of the model are the PA and the shape of the rotation curve,
characterized by a maximum velocity and velocity rise in the central region. 
The procedure will be more fully described in a future paper 
where we apply it to the full MaNGA sample of counter-rotating galaxies \citep{Beom2022}.
The center panels (d and e) of Fig.~\ref{fig:kinematics_fig1a}
show the velocity field of the stellar or gaseous disk
with the kinematic major axes shown as light green lines. 
The PA of the kinematic major axis is also
written in the right bottom corner of each panel.
First, we compared the kinematic major axis of the stellar disk
to the photometric semi-major axis 
which is shown as a green line in the panel for 
the mean surface brightness in g-band.
As seen in the panel (b) and (c) of Fig.~\ref{fig:kinematics_fig1a}, 
the kinematic major axis in the stellar component agrees with 
the photometric major axis of the stellar component; 
the difference between the two is generally within a few degrees, up to 8.4 degrees.
The difference between the kinematic major axes in stellar and gas components
is in the range $\rm 170~to~180~degree$, except for Galaxy C, I and J. 
In Galaxy H and J, the fitted kinematic major axis in the gas disk
is not well constrained due weak emission lines and an irregular gas distribution.
The kinematic major axis in the gas component of Galaxy C
seems not to be properly constrained due to a peculiar feature, 
which is described in the section~\ref{sec:Peculiar_Galaxy_C}.

For a galaxy with a regularly rotating disk and bulge,
the kinematic minor axis would appear as a straight line with zero velocity in the velocity map.
Galaxy G has a straight minor axis 
in the stellar and gaseous velocity fields (d and e panels) 
of Fig.~\ref{fig:kinematics_fig1b}.
On the other hand,
Galaxies C, E, F, I and J show a distorted shape of the kinematic minor axis
in the d or e panels of Fig.~\ref{fig:kinematics_fig1a}. 
In particular, considering that 
the kinematic minor axis appears to be more disturbed
in the gas disk compared to that in the stellar disk,
this might be a result of the perturbation in the kinematics
resulting from the formation of the counter-rotating gas disk.

\subsubsection{Rotation velocities along the major axis}
\label{sec:Rotation_velocities}

We analyzed the velocities along the major axes of the counter-rotating galaxies, as a first order approximation to the rotation curves.
In the (h) panels of Fig.~\ref{fig:kinematics_fig1a},
the velocities are taken from spaxels where the projected distance to the kinematic major axis is within $\rm 1.25~arcsec$. In general, disk galaxies tend to have lower rotational velocities in their stellar than gaseous components due to asymmetric drift. Surprisingly, 
we find weaker rotation of the gas disk by about 
60, 30, and $\rm 90~km~s^{-1}$ compared to 
the stellar disks in galaxies C, E and F, respectively. 
For galaxy E, the gas velocity dispersion is higher and could "make up" for the lower rotational velocity. This is not the case for Galaxies C and F. A slower rotation of the gas disk could arise 
if the disk were not edge-on with respect to the line of sight. 
However, the magnitude of the rotation decrease appears 
to be more severe than that resulting from an uncertainty in the inclination. 
E.g. if inclination were to lower the rotation by about 30 km~s$^{-1}$ from a rotation velocity of 200~km~s$^{-1}$, the gas disk would be inclined by about 60  degrees instead of edge-on. 
Considering the observed spatial distributions of the gas disks, 
this seems unlikely. This weak rotation may indicate 
that for these gaseous counter-rotators, insufficient  
time has passed to reach a kinematic equilibrium configuration.
This implies that the gas disk is not settled yet
after the gas accretion.
This interpretation may also supported by the slight tilt 
of the gas disk plane (five to ten degrees) 
to the stellar disk plane according to the PAs.
Weaker rotation in the gas disks 
is also present Galaxy I and J. 
However, in these cases this effect may be due to the irregular distribution of the emission line gas and the large uncertainty
in the ionized gas velocities due to weak emission.
Galaxy A, B, D, G, and H show similar rotational velocities
in the stellar and gas components.

\subsubsection{Peculiar Feature in the velocity field of Galaxy C}
\label{sec:Peculiar_Galaxy_C}

The velocity field of the ionized gas in Galaxy C shows 
a peculiar feature. The ionized gas in the thick disk of about 2.5 kpc above and below the disk plane
has a receding velocity of about 70~km~s$^{-1}$ with respect to the systemic velocity of the galaxy. This feature remains present at higher S/N cuts in the fitting of the spectra.
It appears with a red color in panel (e) of Fig.~\ref{fig:kinematics_fig1a}. 
This gas seems present on both the approaching and receding side of the galaxy. 
It seems as if receding ionized gas is surrounding the gas distribution. 
The IFU for this object only covers the central part, 
so this is possibly not a whole disk phenomenon but could be concentrated 
around the central region only. 
Further modeling and observations would be needed to confirm 
if this feature might be evidence for gas accretion or related to 
a central outflow. 

\subsubsection{High Velocity Dispersion regions}
\label{sec:High_VD}

We searched for regions with high velocity dispersion.
These might indicate disturbed features remaining after galaxy interaction. 
However, the high inclination makes it difficult to distinguish
if a high velocity dispersion is intrinsic or arises from a
projection of the rotational velocities along line of sight.
In addition, the velocity dispersion values have high uncertainty
for low S/N spaxels, which are found on the outskirts
of the stellar and the gas distribution.
With these considerations, possibly significant high velocity dispersion regions 
are found in the gas disks of Galaxy E, I, and J and the stellar disk of Galaxy B.
As seen in each g panel in Fig.~\ref{fig:kinematics_fig1b}, 
high velocity dispersion regions in the gas disk of Galaxy E, I and J 
appear along the kinematic minor axis.
This is in the agreement with the disturbed feature 
of the kinematic minor axis discussed above.
In panel (f) of Galaxy B in
Fig.~\ref{fig:kinematics_fig1a}, 
we see a high velocity dispersion region near the midplane 
on the right side of the disk.
At the same location in panel (d), 
there appears an somewhat abrupt change in the stellar velocity field, 
shown by the dark blue area.
The abrupt change of the stellar velocity 
and the high velocity dispersion may arisen from the two stellar component rotating in the opposite direction each other. 
However, there is no obvious evidence of 
a second stellar component in the spectrum, 
such as double peaks of the absorption line.

The other galaxies show a typical distribution
of the stellar velocity dispersion as high at the center
and low in the disk. 
In particular, Galaxy H, I and J, the relatively massive counter-rotators,
show noticeable regions of high velocity dispersion 
at the bulge with a round shape consistent
with their high mass and high bulge-to-total ratios 
(0.35, 0.49, 0.47, respectively).

\subsection{Ionized Gas Properties} 
\label{sec:ionized_gas}

\subsubsection{Emission Line Strengths} 
\label{sec:emission_line_strength}

The strength of the emission lines were determined 
on the integrated spectrum of each galaxy.
We summed the spectrum from only the spaxels 
where the mean surface brightness in the g band 
or the peak of the flux in the H$\alpha$ emission wavelength 
is greater than 3.
Due to limited IFU coverage, the measured values can be lower limits;
this would depend on the ionized gas distribution in a galaxy. 
We show the primary sample (IFU coverage $\sim$ $\rm 1.5~R_{eff}$) of MaNGA 
as grey dots and purple polygons
and the secondary sample (IFU coverage $\sim$ $\rm 2.5~R_{eff}$)of MaNGA as black dots and red polygons
in Fig.~\ref{fig:BPT}.
For the counter-rotating galaxies, 
the distribution of ionized gas is concentrated in the center
so that any underestimate due to IFU coverage is likely insignificant. 
However, in general for the edge-on star-forming galaxies, 
the strength of the emission lines could be significantly underestimated
due to the IFU coverage.

\begin{figure}
\centering
\includegraphics[scale=0.45]{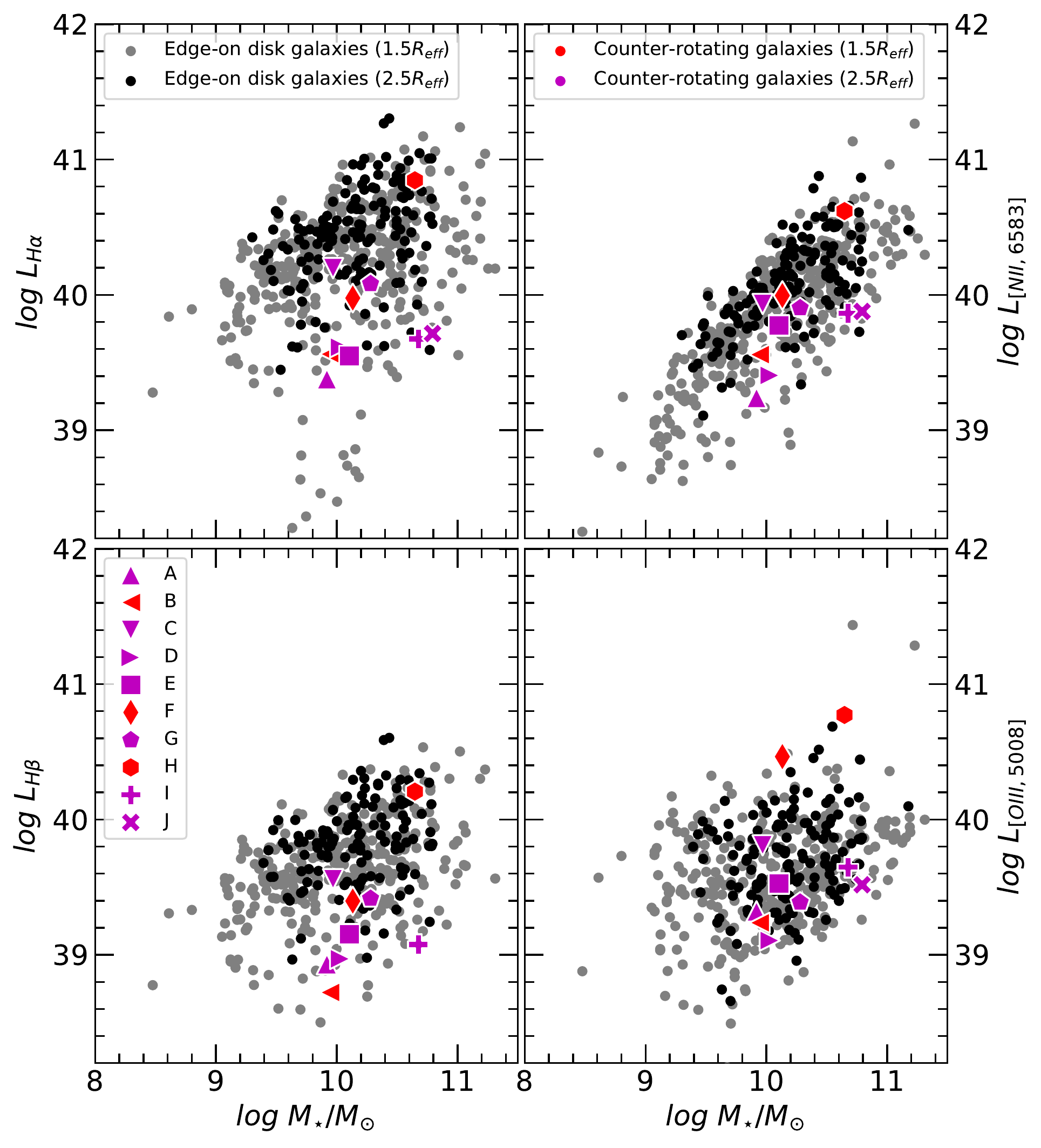}
\caption{The luminosity of 
H$\alpha$, H$\beta$, [NII], and [OIII] emission lines of 
the edge-on galaxies.
The luminosity is in the unit of erg~s$^{-1}$ on a log scale.
Depending on IFU coverage, the edge-on disk galaxies are presented 
as black dots (2.5~R$_{eff}$) and grey dots (1.5~$R_{eff}$). 
The counter-rotating galaxies are presented as red (2.5~$R_{eff}$)  
or purple (1.5~R$_{eff}$) polygons.
Emission line strengths for the galaxies with IFU coverage out to  1.5~R$_{eff}$ (grey dots and purple polygons)
may be lower limits.
Galaxy J is not presented in  the left
bottom panel, because its H$\beta$ is not detected.
\label{fig:elumi}}
\end{figure}

Fig.~\ref{fig:elumi} shows emission line strengths
for the MaNGA edge-on disk galaxies.
As seen in the left panels of Fig.~\ref{fig:elumi}, 
Galaxies C and H show comparable strength of H$\alpha$ and H$\beta$
to that of other edge-on disk galaxies with strong emission lines.
The other counter-rotators have weak balmer lines.
On the other hand, all counter-rotating galaxies 
show strength of [NII] and [OIII] comparable to that in strong emission line galaxies.
Galaxy C and H can be classified as 
strong emission line galaxies. 
Galaxy A, D, I and J would be classified as weak emission line galaxies, 
but not with weak [OIII] emission.
In particular, Galaxy F and H have quite strong [OIII].
Galaxy B, E, F and G have weak Balmer lines, 
but relatively strong [NII] and [OIII].

\subsubsection{Emission Line Ratios} 
\label{sec:emission_line_ratio}

\begin{figure}
\centering
\includegraphics[scale=0.35]{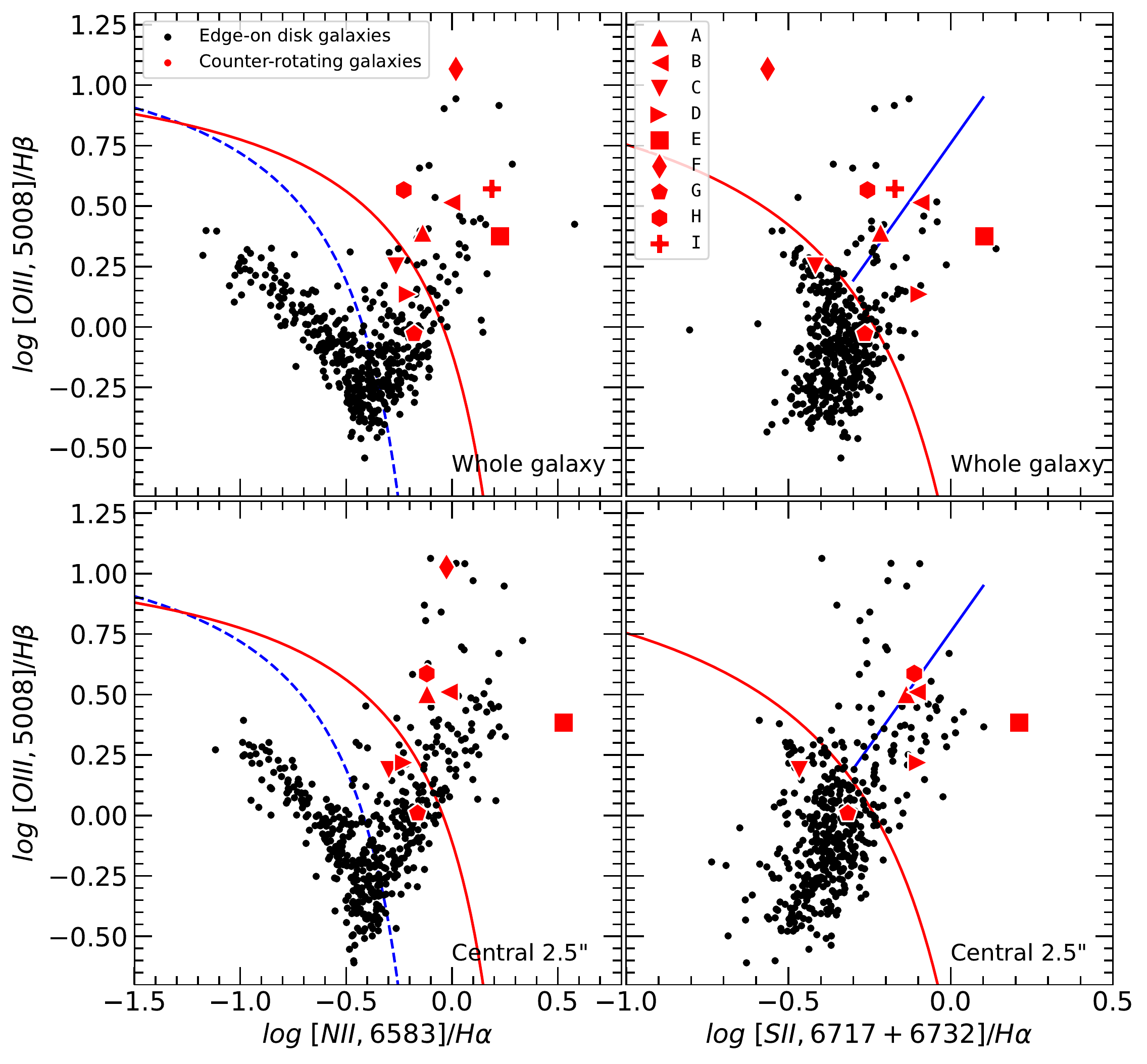}
\caption{The diagnostic diagrams of the counter-rotating galaxies and 
the edge-on comparison sample 
using the emission line ratios 
[NII]/H$\alpha$, [SII]/H$\alpha$ and [OIII]/H$\beta$. 
The counter-rotating galaxies are shown as red polygons and 
the edge-on disk galaxies are shown as black dots.
The lines indicating classifications are taken from \citet{BPTlines}.
To the left of the dotted blue line are
the star forming galaxies.
To the right and above the red line are AGN (above blue line) 
or LINER/LIER (below blue line) galaxies.
The upper panels show the diagnostic diagrams based on 
the integrated spectrum of each galaxy, whereas 
the bottom panels show only the central 2.5~arcsec region of each galaxy.
Galaxy J is not presented in the figure due to non-detection of 
H$\beta$.
Galaxy I is omitted in the bottom two panels due to non-detection of 
H$\beta$ and [SII] in the central 2.5 arcsec spectrum.
The counter-rotating galaxies are located in the AGN/LINER regime 
or in the composite area, not in the star formation locus.
 \label{fig:BPT}}
\end{figure}

The emission line ratios are analyzed with the standard diagnostic diagrams,
[NII]/H$\alpha$ vs. [OIII]/H$\beta$ (the BPT diagram) and 
[SII]/H$\alpha$ vs. [OIII]/H$\beta$ (the VO diagram)
\citep{BPTdiagram,VOdiagram}.
Fig.~\ref{fig:BPT} shows these diagrams with 
the classification boundaries taken from \citet{BPTlines}. 
The upper two panels show the ratios 
measured in the integrated spectrum of
the whole galaxy.
The bottom two panels show the line ratios
determined from the integrated spectra of 
the central 2.5~arecsec regions.
All panels only show galaxies for which all emission lines are detected 
with a SNR greater than 3.

As seen in the panels of Fig.~\ref{fig:BPT},
Galaxy F falls in the AGN galaxy regime.
In order to confirm the possible presence of an AGN,
we searched for a radio detection in the VLASS (VLA Sky Survey)
\citep{VLASS}.
Galaxy F is confirmed as a radio source 
with $S_{total} = 10.217 \pm 0.214~mJy$
at $2 < \nu < 4~GHz$ ($S_{peak}=9.902~mJy~beam^{-1}$), 
which appears as a Single-Gaussian fit source 
in the VLASS catalog \citep{VALSSquick}.
None of the other counter-rotators are detected in the VLASS.

Galaxy A, B, and H  are on the boundary of AGN/LINER galaxies,
and Galaxy D and E are in LINER(low ionization nuclear emission regions) regime. In their spectrum, 
the emission line strengths from the central region are dominant,
so their emission line ratios are similar in the both of 
the entire spectrum and the central spectrum. 
Therefore, we can conclude 
that Galaxies A, B, H, D and E may have a weak AGN or LINER at their center,
probably not a LIER (more extended low ionization emission-line regions) 
classification in which the extended ionized gas would be dominant.
In the central 2.5~arcsec spectrum of Galaxy I,
H$\beta$ and [SII] are not detected and it is not presented in the bottom two panels. 
Considering the emission line ratio estimated 
from the integrated spectrum of the whole disk
and the centrally concentrated distribution of ionized gas,
Galaxy I can plausibly be considered a LINER.
The velocity dispersion of the emission lines can indicate if shock ionization could play a role. 
Seven counter-rotators have overall low velocity dispersion below $\rm 100~km~s^{-1}$, while galaxies E, I, and J shows central regions of high velocity dispersion, possibly indicating shock ionization. Galaxy E in particular is far into the LINER regime in the BPT diagram.

Galaxies C and G are in the composite regime in Fig.~\ref{fig:BPT}.
Similar to the other counter-rotators, 
the ionized gas in the central region is dominant for these galaxies. 
Galaxy J's H$\beta$ emission is too weak to determine its classification on the diagnostic diagrams. Galaxy J does have a tentative outflow feature in its ionized gas distribution in the form of 
a gas blob 
above its stellar disk in H$\alpha$, [NII], and [OIII]
located about 5 arcsec (5.5 kpc) off the center perpendicular 
to the stellar disk. There is no stellar counterpart. 
This region also has high velocity dispersion (above $\rm 150~km~s^{-1}$) at one side. 
A deeper observation would help to confirm the existence of the outflow and the possible presence of an AGN. Overall, it is striking that none of the counter-rotators are in the pure star formation location on all line diagnostic diagrams.

\subsubsection{Gas Phase Metallicity} \label{sec:gas_metallicity}

The gas phase metallicity may shed light on the nature 
of the counter-rotating gas.
We estimated the gas phase metallicity using the indicator O3N3 which is the ratio of [OIII]/H$\beta$ to [NII]/H$\alpha$.
This gas phase metallicity indicator is 
considered a better indicator for spectra including
diffuse ionized gas (DIG) and HII regions \citep{Curti2017}.
Due to the centrally concentrated emission line distribution of the ionized gas in the counter-rotating galaxies, the gas phase metallicity we derive is mostly determined by 
this centrally ionized gas.
The gas phase metallicity is listed in Table~\ref{tab:photometric}.
Most of the counter-rotating galaxies 
have a metallicity in the range of 8.58 to 8.72 in 12+log(O/H), which is within about $\pm 0.1$ from the solar metallicity (12+log(O/H)$_{\odot}$=8.69).
Galaxy F has a metallicity of 8.51, a little bit lower than others.
However, none of the gas phase metallicities are so low 
as to support {\it recent} accretion from the IGM or merging with a metal-poor gas low mass dwarf galaxy. A radial metallicity gradient 
is difficult to measure for the counter-rotators 
due to lack of ionized gas in the outer disk.
A meaningful gradient can be measured only in Galaxy G and H,
which is flat. 

\subsection{Star Formation Rates} \label{sec:SFR}

We estimated the star formation rate (SFR) using the
H$\alpha$ luminosity
and the infrared luminosity from WISE.
We estimated the SFR from a ``mixture method" which combines the H$\alpha$ luminosity with the 24 $\micron$ IR luminosity \citep{Calzetti2007}.
In this method, we adapted the equation of \citet{Murphy2011} 
with an `$\alpha$' coefficient of 0.042 which was suggested as a better value 
for edge-on spiral galaxies \citep{Vargas2018}.
For the 24 $\micron$ IR luminosity, we converted 
the infrared luminosity in W4 band (22 $\micron$) 
into 24 $\micron$ with a conversion factor of 1.03.
This conversion factor was 
estimated by the tight correlation of luminosity between two bands \citep{Wiegert2015,Jarrett2013}. 
The equations are:

\begin{equation} 
\centering
\begin{split}
SFR_{mix}=5.37\times10^{-42} L(H\alpha_{corr}) \\
L(H\alpha_{corr})=L(H\alpha_{obs})+\alpha \cdot \nu L_{\nu}(24\micron) \\
L_{\nu}(24\micron)=1.03 \times L_{\nu}(22\micron,W4) 
\end{split}
\label{SFR}
\end{equation}
where L(H$\alpha$) and $\nu L_{\nu}$ values are in the unit of 
$\rm erg~s^{-1}$, and $\rm L_{\nu}(22,24\micron)$ is in 
$\rm erg~s^{-1}~Hz^{-1}$.

\begin{figure}
\centering
\includegraphics[scale=0.52]{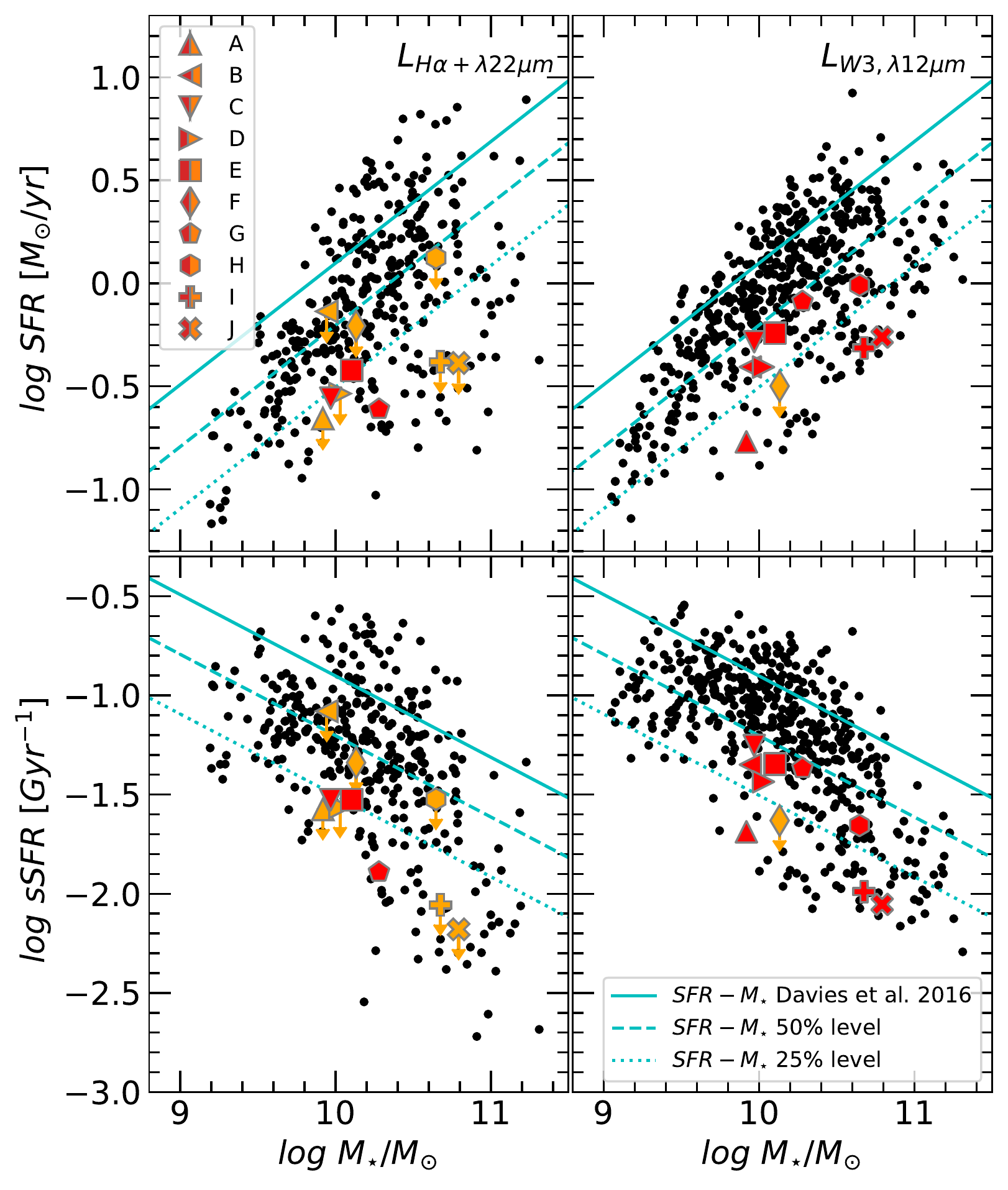}
\caption{The star formation rates (SFR) and the specific star formation rate (sSFR)
of the counter-rotating galaxies and 
the edge-on comparison sample. 
The left panels show SFR and sSFR evaluated with the ``mixture method", a combination of the H$\alpha$ plus W4 luminosity.
The right panels show SFR and sSFR estimated from the WISE W3 luminosity.
The bottom two panels show sSFR, i.e. the SFR divided by the stellar mass.
The counter-rotators are shown as red (detected in WISE) and yellow (upper limits in WISE) polygons.
For the comparison sample (black dots), 
we do not include galaxies having only an upper limit to the WISE magnitude.
The solid line indicates the star formation main sequence 
calculated by the W3 luminosity of \citet{Davies2016}.
The dashed and dotted lines indicate 50\% and 25\% levels of 
the star formation rates of the main sequence lines. 
The gaseous counter-rotating galaxies have SFRs lower than the 50\% level of 
the star formation main sequence.
 \label{fig:SFR}}
\end{figure}

\begin{figure*}
\centering
\includegraphics[scale=0.5]{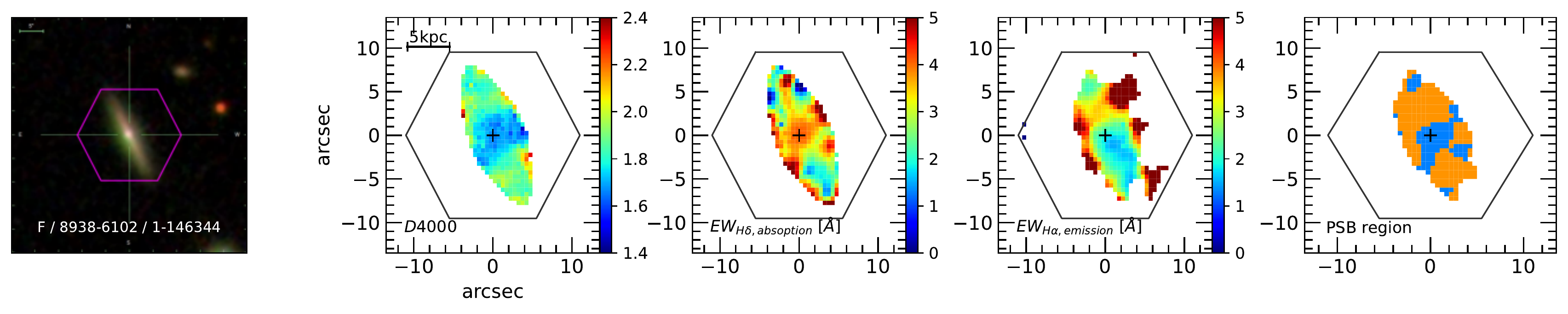}
\caption{A post starburst signature for galaxy F. 
The SDSS broad band image, spectral indices D4000 and the equivalent width of H$\delta$ absorption
and H$\alpha$ emission line, from left to right.
The rightmost panel shows a post starburst region in the central region of Galaxy F 
as the light blue colored area with the overall stellar distribution (orange). 
\label{fig:PSBR}}
\end{figure*}

Since not all galaxies were detected in W4 by WISE, 
we also evaluated the star formation rates 
using the W3 magnitude as an alternative method, following \citet{Davies2016}:

\begin{equation} 
\centering
log SFR_{W3}=0.66\times(log L_{\nu}(W3)-22.25)+0.16
\label{SFR_w3}
\end{equation}
with $\rm L_{\nu}(W3)$ in $\rm W~Hz^{-1}$, not $\rm erg~s^{-1}~Hz^{-1}$. 
The mixture method is likely more reliable 
to estimate the star formation rate than the W3 luminosity method
because W3 is also sensitive to emission from polycyclic aromatic hydrocarbon particles (PAHs).
However, since the H$\alpha$ emission strength may be limited by the IFU coverage, the W3 method provides a useful alternative.
In the figures we exclude the comparison edge-on galaxies that only
have upper limits in both of $W3$ and $W4$.
Among the counter-rotators, galaxy F galaxy only has an upper limit in both W3 and W4
and its star formation rate is presented an an upper limit value.

Fig.~\ref{fig:SFR} shows the star formation rates (SFR) 
and the specific star formation rates (sSFR) of edge-on galaxies 
with the star formation main sequence lines for nearby galaxies \citep{Davies2016}.
The sSFR is the ratio of the star formation rate to the stellar mass.
Galaxies B, D, E, F, and H have SFRs in the range 
25 to 50\% of the level of the SFR of the star formation main sequence, and 
Galaxies A, C, G, I, and J have lower SFRs than the 25\% percentile.
In particular, Galaxy I and J have sSFR less than $10^{-11}~yr^{-1}$,
which is approaching the range of values for quiescent galaxies.
The other counter-rotating galaxies
show low specific star formation rates (sSFR) 
at the level of $10^{-10}$ to $\rm 10^{-11}~yr^{-1}$, 
consistent with being in the the transition stage 
between star forming galaxies and quiescent galaxies.
We also compared the star formation rates 
with four different star formation main sequences 
obtained by various methods taken from previous studies \citep{Davies2016,Jarrett2017}.
In all these comparisons 
the counter-rotating galaxies show low star formation rates.

\subsection{Search for Post Starburst Signature} \label{sec:PSBR}

It is likely that the acquisition of the counter-rotating gas disk involves transport of gas to the central region of the galaxy, for example through loss of angular momentum if there were a pre-existing co-rotating gas disk present. Such concentration of gas might lead to star formation bursts in the central regions. 
Since current SFRs are low, we searched for a post star burst spectroscopic signature 
in the central regions of the counter-rotators which would be present up to several hundreds Myrs after a starburst.
Following \citet{Chen2019}, a post starburst region is indicated by 
EW(H$\delta$) > 3~\AA~ in stellar absorption, 
EW(H$\alpha$) < 10~\AA~ in emission,
and $\rm log~EW(H\alpha) < 0.23 \times EW(H\delta) - 0.46$.
Note that the equivalent widths of both of the H$\delta$ {\it absorption} 
and the H$\alpha$ {\it emission} lines
are defined to be positive with these criteria. 
With these criteria we only find a potential post star burst region in Galaxy F.
Fig.~\ref{fig:PSBR} shows the D4000 and H$\delta$ absorption line spectral indices and the equivalent width of H$\alpha$ emission line.
The implied post starburst region according to the above criteria,
is the green region in the most right panel. 
It extends about 4 kpc across at the center,
whose age is about 150 to 300 Myr with the method of \citet{Chen2019}.
Since we expect past massive star formation in counter-rotators, 
the fact that only one counter-rotator has the PSB region at the center
may be surprising. 
We will revisit this in the discussion (the section \ref{sec:discu}).

\begin{figure*}
\centering
\includegraphics[scale=0.6]{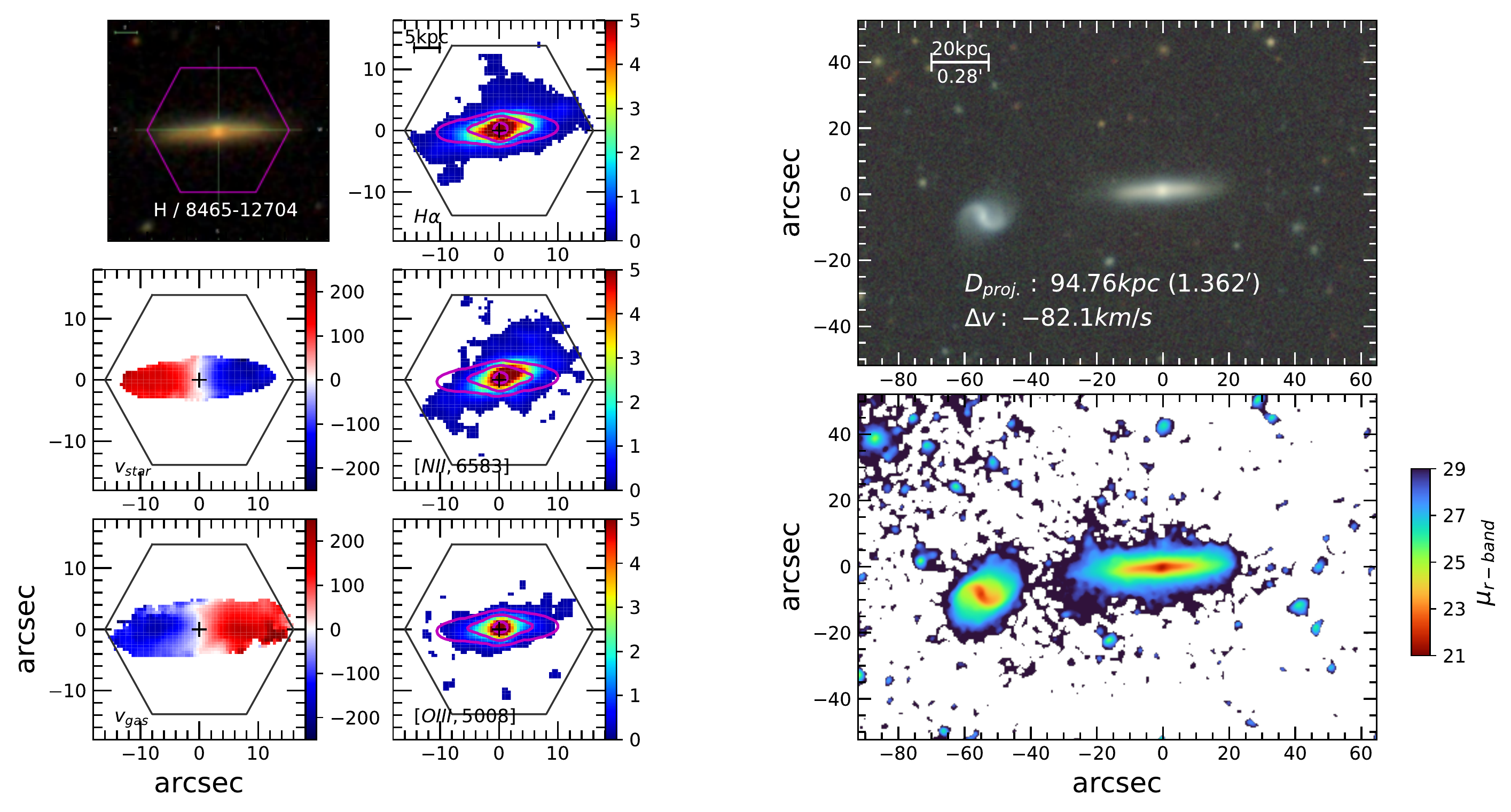}
\caption{SDSS broad band image with IFU outline,
velocity fields of stellar and gas disks, ionized gas distributions,
and a deeper optical broad band image of Galaxy H and its close companion.
The line of sight velocity is in units of km~s$^{-1}$, 
In the second column panels, 
the strength of the H$\alpha$, [NII], and [OIII] emission lines
is presented in units of $\rm 10^{-17}~erg~s^{-1}~cm^{-2}~spaxel^{-1}$
with purple contours showing the stellar distribution 
as the surface brightness in the g band.
The rightmost two panels show a DECaLS color image from g, r, and z bands and
an APO 3.5m r band image with an exposure time of 3600~s.
The APO image is smoothed by a median filter of 11 by 11 pixels.
\label{fig:disks_8465}}
\end{figure*}

\subsection{Search for Evidence of Ongoing or Post Merger Signatures} \label{sec:interaction}

Deep images of all counter-rotating galaxies 
were taken using the ARC 3.5m telescope at APO.
In this section we focus on searching for ongoing or post merging signatures,
such as a lopsided disk in the outskirt or a tail structure.

We found ongoing merger signature on Galaxy H and its nearby companion.
The companion galaxy is at a projected distance of 95~kpc
and a velocity difference of 82~km/s, which is close enough 
to make it a potential interacting pair. 
There are now several pieces of evidences supporting 
on-going galaxy interaction, see Fig.~\ref{fig:disks_8465}. 
Its ionized gas distribution (colored maps) is somewhat tilted 
relative to the stellar disk (purple contours). 
The direction of the tilted disk is in the direction towards the companion. 
In addition, faint extended continuum light 
is present between the two galaxies 
in Fig.~\ref{fig:disks_8465} taken by APO 3.5m telescope.
The right bottom panel shows the APO image smoothed by 
a median filter of 11 by 11 pixels. 
The asymmetric shapes of the outskirt of the disks, 
are at about 28.0 to $\rm 28.5~mag~arcsec^{-2}$ in r-band. 
This may imply a faint structure connecting between the two galaxies 
or a lopsided disk in Galaxy H.
The companion galaxy is a late type spiral galaxy (SDSS J131239.46+482151.8).
Its inclination is close to face-on so that we can clearly see 
two strong spiral arms in the disk.
The absolute magnitude is -19.6~mag in r band and the stellar mass is $\rm 6.635 \times 10^9~M_{\odot}$ $(log~\rm M_{\star}/M_{\odot}=9.822)$, which corresponds to 
a stellar mass ratio to galaxy H of $0.15$ $(1:6.73)$. 
While lower in mass than Galaxy H, it is more massive than a dwarf galaxy. 
The companion has a blue color ($0.413$ in g-r and $2.03$ in NUV-r).
Based on its SDSS spectrum (SDSS specobjid: 1442302565864728576), 
it is a strong emission line galaxy with emission line ratios 
that place it in the star forming region on the BPT diagram.
Its gas phase metallicity is 8.73, which is higher by 0.22 dex 
compared to 8.51 of Galaxy H.
With the other characteristics addressed in the previous sections, 
this ongoing merger signature needs a further discussion
in order to understand the formation of a counter-rotating galaxy 
by merging with a gas rich companion galaxy. 
We will visit this again in the section \ref{sec:origin}.

\begin{figure}
\centering
\includegraphics[scale=0.43]{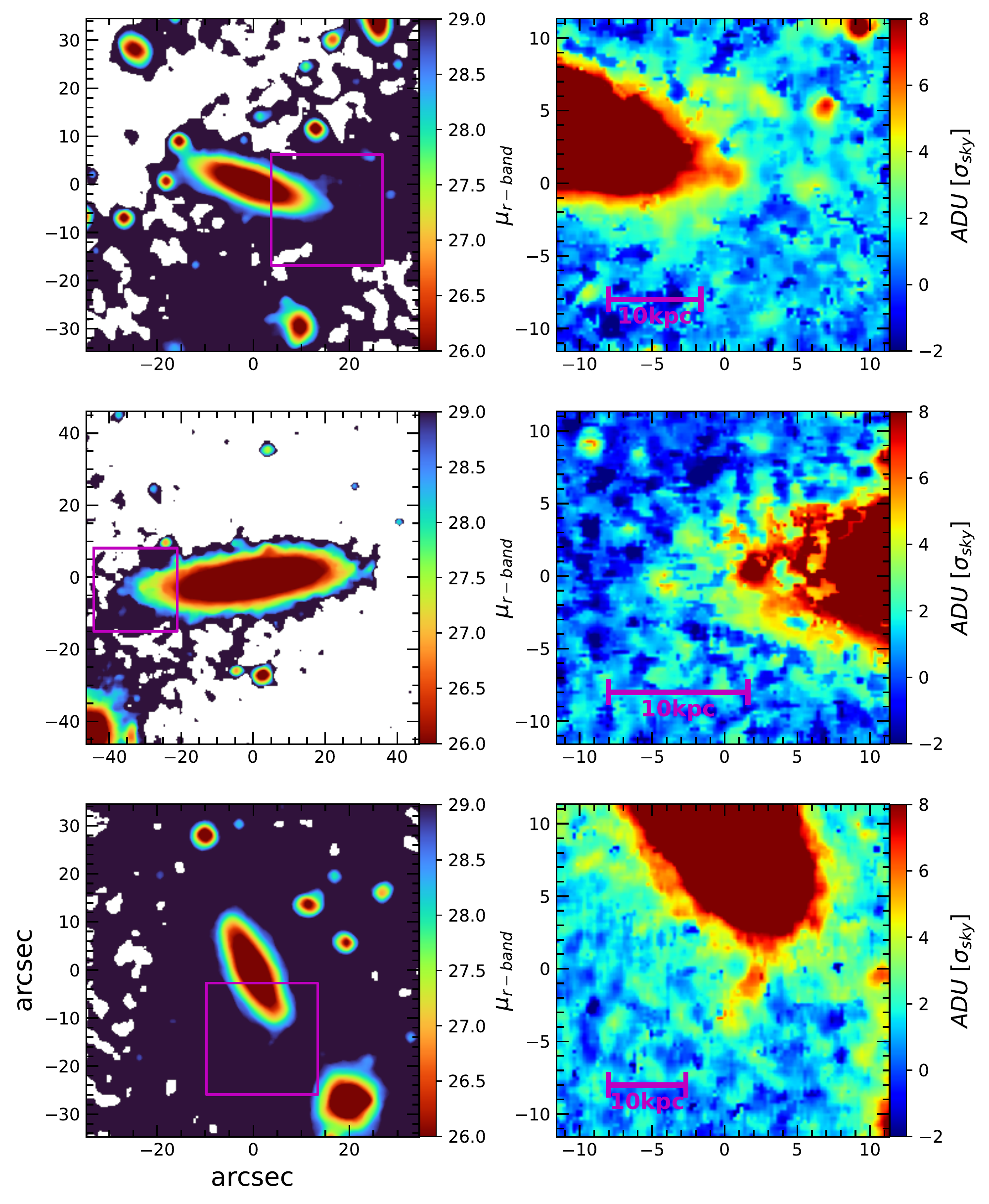}
\caption{Deep image in r band  
observed  with the ARC 3.5m telescope for Galaxy B, E, and F (from top to bottom panels), the only objects to show faint features of interest in the outer edges of the their disks in our APO 3.5-m deep images. 
The left panels are the r-band images smoothed by an 11 by 11 median filter,
shown as the surface magnitude. 
The right panels are zoomed-in images of the region shown 
by the purple square in the left panels. 
These show potential features of interest, on a linear scale 
in the unit of the noise (standard deviation) of 
the sky of each image in observed counts (ADUs). 
 \label{fig:deepimage}}
\end{figure}

We also found the post merger signatures on the broad band deep images.
Fig.~\ref{fig:deepimage} shows the deep image of Galaxy B, E and F
in surface magnitude scale (left panels) and linear scale (right panels).
The left panels are presented as
the images smoothed by the median filter using 11 by 11 pixels 
in order to show the extended feature. 
The right panels show a zoom-in image in linear scale not smoothed 
in order to show how bright features are compared to the sky noise
of each image.
As seen in the first row panels of Fig~\ref{fig:deepimage}, 
Galaxy B has faint small features on the right outskirt of the disk. 
A tail structure at the position of (0,0) in the right panel
appears as 5 to 6 sigma level of the sky noise. 
This tail feature is located at the edge of the stellar disk
on the semi-major axis direction.
Also, a faint feature at the position of (2,5) in the right panel
appears as blobs, which is relatively bright compared to surroundings
and extended on about 3 by 7~kpc region.
This feature has about 4 sigma level of the sky noise.
Note that a bright round shape feature on the position of (7,6)
in the right panel is identified as a faint star. 
In the middle two panels, Galaxy E had lopsided outskirt on the left side 
of the stellar disk.
In the smoothed image of the left middle panel, 
we can see lopsided stellar distribution on the left outskirt of the disk,
which appears as about $\rm 28.5~mag~arcsec^{-2}$.
In the zoom-in image on the right middle panel,
the lopsided feature is made by some extended features
of 4 to 7 sigma level of the sky noise.
Its extent is about 5 by 7~kpc.  
In the bottom two panels, 
we can see a tail feature in the southern part.
This structure is about 5~kpc in length, 
and its brightness is a factor of 4 to 6 sigma of the sky noise.
Note that all features discussed in this section are faint in the range of 
28.0 to $\rm 28.5~mag~arcsec^{-2}$, which should be cautiously accepted. 
Although we discussed only features widely extended to several kpc and 
brighter than 4 to 6 sigma level of the sky noise, 
we can not completely rule out a possibility that 
a feature consists of two to several faint foreground stars or 
a coincident feature is appeared by uneven sky.

\section{Discussion} \label{sec:discu}

\subsection{The Frequency of Counter-rotating Galaxies} \label{sec:frequency}

We found ten gas counter-rotating galaxies among 523 edge-on disk galaxies 
in MaNGA (MPL-11), a fraction of $1.9 \pm 0.6\%$.
This fraction refers to counter-rotators with co-planar or very close to co-planar gas and stellar disks, and with gas disks of comparable size to the stellar disks. 
In the era before IFU surveys,
\citet{KannappanFabricant2001} found 4 gaseous counter-rotators among 55 galaxies of 
various morphological types (2 out of 3 ellipticals, 
2 out of 14 S0 galaxies, and none among 38 late type galaxies).
Hence they identified 2 gaseous counter-rotators out of 52 disk galaxies.
\citet{Pizzella2004} also found 2 gaseous counter-rotators 
in a sample of 50 S0 and spiral galaxies.
The fraction of 4\% found in both studies 
is higher than we find although it agrees within the uncertainties.

For recent, much larger IFU surveys, \citet{Bryant2019} using the SAMI data and \citet{Li2021} using the MaNGA sample find fractions of about 3\% gaseous counter-rotators among galaxies where kinematic position angles could be fitted 
for the stellar and gaseous components. This too 
is higher than our value of 1.9\%.
Those samples, however, included elliptical galaxies and as previous papers have shown, the fractions of gaseous counter-rotators are higher for elliptical and S0 galaxies,
e.g. 23.5\% from E/S0 in \citet{KannappanFabricant2001}, and 
20\% and 24~\% for S0 galaxies in \citet{Bertola1992,Kuijken1996}, respectively. High fractions of misaligned, i.e. not restricted to counter-rotating co-planar disks but in the range 30<$\Delta$ PA<150 degrees
have also been reported in observational and simulation studies
as a common kinematic characteristic for early type galaxies
\citep{Pizzella2004,Davis2011,Bryant2019,Duckworth2019,Duckworth2020,Khim2020,Xu2022}.

We can estimate an upper limit to the fraction of gaseous counter-rotators for the S0 class in our sample. 
The fraction of S0 galaxies in a group environment 
according to \citet{van_den_Bergh2009} 
classifying the Shapley Ames catalog, 
is about 20\% among disk galaxies.
Applying this to our edge-on disk galaxy sample of 523
and assuming that all the gaseous counter-rotators in this study are S0,
the fraction of the gaseous counter-rotators within S0 galaxies 
would be at most 10~\%.

In summary, for our edge-on sample we find a fraction of 1.9\% counter-rotating disk galaxies, somewhat lower than found in previous work. The difference may be due to our sample selection which was restricted to 
undisturbed appearing disk galaxies with spatially extended gas disks and close to co-planar gas and stellar disks.

\subsection{The Environments} \label{sec:environment}

We investigated the environments of the counter-rotating galaxies 
using spectroscopic redshifts and optical images from SDSS.
Potential group members of the counter-rotators are taken to have
a velocity difference of $\rm \pm 500~km~s^{-1}$ 
and a projected distance on the sky within 600~kpc.
We also searched a 16' by 16' field around the targets
to identify potential group members without a redshift.
We verified that our results agree with other studies of MaNGA target environments, including
\citet{Tempel2017} and GEMA VAC (Galaxy Environment for MaNGA, Value Added Catalog. 
The method of GEMA is described in \citet{Argudo2015}, 
and the value added catalogs of MaNGA are described in \citet{mangadr17}.
We summarize the findings in Table~\ref{tab:photometric}. 
Galaxies C and D appear to be isolated field galaxies.
The other galaxies are in groups with 1 to 4 additional members. 
This implies that the counter-rotators are typically found in small and loose groups. 

We point out a few notable environmental facts here.
First, Galaxy G is located in a rather crowded region on the sky.
Within a 4 by $\rm 3~arcmin^{2}$ field,
it we find 5 other galaxies, 2 high redshift galaxies,
($\rm z\sim0.072$), and 7 foreground stars.
The 3 group members of galaxy G with a confirmed spectroscopic redshift are relatively far away from it (351, 468, 570~kpc, respectively.)
The 5 nearby galaxies do not have spectroscopic redshifts, 
but three of these galaxies are similar in angular size to galaxy G. 
Hence Galaxy G could have more group members or nearby companion(s) 
than indicated in the table.
Secondly, Galaxy B, E, F, H and I have a nearby galaxy among group members 
with a projected distance less than $\rm 150~kpc$.
Of these, only Galaxy H shows some evidence 
for an on-going interaction between the counter-rotator and its companion galaxy 
as discussed in the section~\ref{sec:interaction}.
Thirdly, Galaxy J has an interesting group member MRK 273
which is at 5.476~arcmin (255~kpc) in projected distance and 21~$km~s^{-1}$ radial velocity difference.
MRK 273 is well known an on-going interacting galaxy, 
which is also known as a starburst, ULIRG, and Megamaser galaxy \citep{Mrk273megamaser,Mrk273AGN}.

\subsection{Characteristics of the Counter-rotating Galaxies 
and their Formation} \label{sec:conclusion}

A gaseous counter-rotator has to be formed through gas accretion 
in retro-grade direction. 
Its formation process may be expected to result in the angular momentum loss in the gas component, if any pre-existing co-rating gas is present, leading 
to a centrally concentrated distribution of the gas component.
Subsequent results would include
star formation at the center, and possibly feeding an AGN in the presence of a nuclear black hole.
In this section we summarize how the characteristics derived in this paper
may be interpreted in the formation processes 
of gaseous counter-rotating galaxies.

\subsubsection{Angular Momentum Loss} \label{sec:discussion_AM}

Angular momentum loss is not only an important characteristics itself
but also the primary driver for subsequent processes.
Since the angular momentum loss results from hydrodynamical friction between any pre-existing gas and the infalling gas,
the amount of pre-existing gas is crucial to 
the angular momentum loss and the time needed for infalling gas to settle down 
to the stellar disk.
The time a misalignment angle 
of the gas component survives has been discussed in simulations \citep{vandeVoort2015,Khim2021}). Our data provide direct evidence for angular momentum loss. As shown in section \ref{sec:Kinematics},
the rotation of the ionized gas in Galaxy C, E and F 
is slower than the rotation of the stellar disk. 
This is opposite to what is typical in a disk galaxy. 
We also find centrally concentrated gas distributions in the counter-rotators.
As discussed in the section \ref{sec:distribution}, 
no strong ionized gas blob is identified in the disk. 
In particular, Fig.~\ref{fig:X2Y2} shows 
the radial concentration of the ionized gas in the counter-rotators.
Secondly, counter-rotators do not have strong dust lanes in their disks (section \ref{sec:photometric}), 
low W2-W3 color (the section \ref{sec:Dust}), 
low emission line strengths (Fig.~\ref{fig:elumi})
and low star formation (the section \ref{sec:SFR}). These data all suggest overall low gas and dust content.

A gas-poor progenitor galaxy might more likely to lead to
a surviving counter-rotating gas disk
since a smaller amount of infalling gas could still sweep the pre-existing gas.
With no data on the pre-existing gas, it is difficult to discuss if the gas concentration in the counter-rotating disk 
is the result of the initial condition or the formation process. 
Interestingly, we also find small radii for the starlight distribution. This may indicate that the inflow of gas has led to enhanced central star formation. However, we do not detect counter-rotating stars and only one galaxy shows a post starburst signature. Thus we cannot exclude that the galaxies were small to begin with.

\subsubsection{Star Formation Signature and AGN} \label{sec:discussion_SF}

The gas concentrated at the center 
can be consumed by the star formation and feeding an AGN.
A young counter-rotator may show high starformation,
observable through a high H$\alpha$ luminosity.
As the counter-rotator ages and star formation ceases, the spectrum would evolve to a PSB (Post Star-Burst) signature.
The final stage might be a quiescent galaxy with a surviving counter-rotating relatively gas poor disk.
In addition, as a result of this star formation, 
a more significant bulge may develop in the gaseous counter-rotators.
In this study, none of the counter-rotating galaxies is confirmed 
as a young counter-rotator with high star formation signature.
We note that the two most massive counter-rotators, 
Galaxy I and J, have specific star formation rate less than
$\rm 10^{-11}~yr^{-1}$, which is close the range of quiescent galaxies. 
The other counter-rotators show low star formation rates, 
but their star formation rate is not low enough 
to consider them quiescent galaxies. 
They seem to be in the intermediate stage 
from the second to the third stages. 
A bulge is  confirmed for all the counter-rotators
as the small 50\% and 90\% light radii, and 
the concentration of the stellar distribution 
(Fig.~\ref{fig:radius} and \ref{fig:X2Y2}).

Galaxy F is the only galaxy confirmed to have a PSB region at its center. 
Galaxy F is also the bluest galaxy in our sample in Fig.~\ref{fig:CMD}.
Moreover, Galaxy F is confirmed as a likely AGN galaxy,
with a radio point source in VLASS
as well as emission line ratios placing it in the AGN section on the diagnostic diagram 
(see details in the section \ref{sec:emission_line_ratio}).
These characteristics can be interpreted as Galaxy F is 
a relatively young gaseous counter-rotator with star formation activity in a few hundreds Myrs and on-going AGN feeding. All this suggests that Galaxy F is the youngest counter-rotator in our sample.

\subsubsection{Formation Origin} \label{sec:origin}

Any formation scenarios involves likely outside origins for the counter-rotating gas component given the angular momentum of the gas disk significantly different to 
that of the stellar disk.
An outside origin of gas would involve gas accretion from the circumgalactic medium or merging with another galaxy.
For the accretion origin, the infalling gas had to have initial angular momentum.
The infall can be both `episodic' or `continuous'
\citep{ThakarRyden1996,Thakar1997}.
For the merging scenario, the merger has to involve a galaxy with enough gas that
the accreted gas can sweep the pre-existing gas of the host galaxy. 
A gas-rich dwarf galaxy is a likely candidate. Past merger scenarios predicted
that major merger would produce an elliptical galaxy \citep{HernquistBernes1991}. 
However, \citet{Zeng2021} showed that a major merger with a spiral-in falling
can make a disk galaxy.  Strictly speaking then we cannot rule out the possibility of a major merger as an origin. However, none of our spectra show any clear evidence for a counter-rotating stellar component and a merger with a dwarf seems more likely. 
The metallicity of the counter-rotating gas could also provide some information.
In the context of the results of \citet{Kumari2021}, 
one may expect low gas phase metallicity  
in the gas accretion origin.
However, low gas metallicity could also result from a merger with 
a dwarf galaxy with low metallicity, such as an SMC-like object.
In this sense it is perhaps surprising that none of the counter-rotators in this paper shows low gas phase metallicity. This suggests either than enrichment of the gas has already happened since the acquisition or that the acquired gas was not very metal poor to begin with. In the accretion scenario this points more to a circumgalactic rather than a intergalactic origin. In the merging scenario, it would point to a more massive dwarf, maybe closer to an LMC type object. However, if enrichment happened after the gas was acquired of if the merger was with a more massive dwarf, one would expect that counter-rotating stars ought to be present and a placing limits on the fraction of such stars would be useful.

Galaxy H is only the galaxy for which we have a stronger hint  
at the formation origin of the counter-rotating gas disk. 
As discussed in the section \ref{sec:interaction},
Galaxy H seems to be undergoing the galaxy interaction with its nearby companion.
It has a nearby companion galaxy at 
95~kpc in the projected distance and 82~$km~s^{-1}$ in the radial velocity difference. 
Considering the gas disk of Galaxy H is tilted toward the the companion galaxy
and the lopsided shape of the outskirt of their disks (Fig.~\ref{fig:disks_8465}), 
Galaxy H and the companion are likely an interacting pair. 
The companion galaxy is a face-on late type galaxy with two spiral arms.
Its stellar mass is 6.73 times smaller than that of Galaxy H. 
Based on the SDSS central spectrum, it is a star forming galaxy.
These physical properties seem appropriate in the light of a possible
minor merger origin of the gas rich companion with galaxy H to produced
the counter-rotating gas disk. The relative blue shift of the companion compared to galaxy H is in the direction of the rotation of the counter-rotating disk on this side.
In particular, strong emission strengths and 
less concentrated distribution of the gas disk of Galaxy H 
can be understood as features of a young gas disk. 
However, other characteristics of Galaxy H 
is contradict to this perspective.
Galaxy H has already red color, small radius of stellar disk,
and  low star formation rate, which are 
features expected in an old counter-rotator.
In particular, the gas phase metallicity of Galaxy H 
is 8.51 lower by 0.2 dex than for the companion, which has 8.76. 
In addition, the spiral arms of the companion has some asymmetry but does not seem so disturbed as might be seen in an interacting galaxy 
whose outer gas disk has been transfered to another galaxy (Fig.~\ref{fig:disks_8465}).
The distribution and the kinematics of the neutral hydrogen gas 
can show the kinematic relation between counter-rotator and its companion
\citep{BureauChung,Chung2012}.
Therefore, further observations of HI gas distribution and its kinematics
in the companion and overall system would be needed to confirm this.

\section{Summary}

We discovered 10 edge-on galaxies with a counter-rotating gas disk of equal size to the stellar disk in a sample of 513 edge-on galaxies in the MaNGA IFU survey. The counter-rotators have relatively small disks in stars and gas, low star formation rates, and low dust content. We detect a potential AGN in one galaxy, and find evidence for a poststarburst population in another one. From deep imaging at APO with the 3.5m we find that a few galaxies show faint distortions in the outer isophotes possibly evidence of past interaction. One galaxy has a close gas rich companion with a blueshift similar to the counter-rotating disk side, but it lacks strong evidence of interaction in its stellar disk. The compact sizes and other properties of the galaxies support the view that transport of gas to the central regions, as might be expected from the angular momentum loss, has occurred. Several galaxies show slower rotation in the gas disk than in the stellar disk. The counter-rotating gas is not particularly metal poor. We cannot distinguish between merger or gas accretion as the origin of the counter-rotating gas. We do find that the counter-rotating galaxies are preferentially found in small group environments and may be on the evolutionary path through the green valley to the quiescent phase.

\section*{Acknowledgements}

We thank the referee for feedback that helped to improve several aspects of the paper.

MB and RAMW acknowledge partial support for this work through NSF grant No. AST-1615594. DB is partly supported by RSCF grant 19-12-00145. YC acknowledges support from the National Key R\&D Program of China (No. 2017YFA0402700), the National Natural Science Foundation of China (NSFC grants 11573013, 11733002, 11922302), the China Manned Space Project with NO.CMS-CSST-2021-A05.

SDSS-IV is managed by the Astrophysical Research Consortium for the Participating Institutions of the SDSS Collaboration including the Brazilian Participation Group, the Carnegie Institution for Science, Carnegie Mellon University, the Chilean Participation Group, the French Participation Group, Harvard-Smithsonian Center for Astrophysics, Instituto de Astrof\'isica de Canarias, The Johns Hopkins University, Kavli Institute for the Physics and Mathematics of the Universe (IPMU) / University of Tokyo, Lawrence Berkeley National Laboratory, Leibniz Institut f\"ur Astrophysik Potsdam (AIP), Max-Planck-Institut f\"ur Astronomie (MPIA Heidelberg), Max-Planck-Institut f\"ur Astrophysik (MPA Garching), Max-Planck-Institut f\"ur Extraterrestrische Physik (MPE), National Astronomical Observatories of China, New Mexico State University, New York University, University of Notre Dame, Observat\'ario Nacional / MCTI, The Ohio State University, Pennsylvania State University, Shanghai Astronomical Observatory, United Kingdom Participation Group, Universidad Nacional Aut\'onoma de M\'exico, University of Arizona, University of Colorado Boulder, University of Oxford, University of Portsmouth, University of Utah, University of Virginia, University of Washington, University of Wisconsin, Vanderbilt University, and Yale University.

This study is partially based on observations obtained with
the Apache Point Observatory 3.5-meter telescope, which is
owned and operated by the Astrophysical Research Consortium. 

This work makes use of HI-MaNGA data taken under proposal IDs: GBT16A-095, GBT17A-012 or GBT19A-127. The Green Bank Observatory is a facility of the National Science Foundation operated under cooperative agreement by Associated Universities, Inc. 
This work also makes use of the ALFALFA survey, based on observations made with the Arecibo Observatory. The Arecibo Observatory is operated by SRI International under a cooperative agreement with the National Science Foundation (AST-1100968), and in alliance with Ana G. Me\'nndez-Universidad Metropolitana, and the Universities Space Research Association.

\section*{Data Availability}

The data cubes of MaNGA final data release (DR17) are publicly available, 
as well as the kinematic maps produced by the MaNGA DAP (Data Analysis Pipeline), 
and can be found at https://www.sdss.org/dr17/manga/. 
The broad deep images taken by APO 3.5m telescope will be shared 
on reasonable request to the corresponding author.




\begin{thebibliography}{}


\bibitem[\protect\citeauthoryear{Abdurro'uf et al.}{2022}]{mangadr17} Abdurro'uf, Accetta K., Aerts C., Silva Aguirre V., Ahumada R., Ajgaonkar N., Filiz Ak N., et al., 2022, ApJS, 259, 35. doi:10.3847/1538-4365/ac4414
\bibitem[\protect\citeauthoryear{Aguado et al.}{2019}]{mangadr15} Aguado D.~S., Ahumada R., Almeida A., Anderson S.~F., Andrews B.~H., Anguiano B., Aquino Ort{\'\i}z E., et al., 2019, ApJS, 240, 23. doi:10.3847/1538-4365/aaf651
\bibitem[\protect\citeauthoryear{Ahumada et al.}{2020}]{DR16} Ahumada R., Prieto C.~A., Almeida A., Anders F., Anderson S.~F., Andrews B.~H., Anguiano B., et al., 2020, ApJS, 249, 3. doi:10.3847/1538-4365/ab929e
\bibitem[\protect\citeauthoryear{Argudo-Fern{\'a}ndez et al.}{2015}]{Argudo2015} Argudo-Fern{\'a}ndez M., Verley S., Bergond G., Duarte Puertas S., Ramos Carmona E., Sabater J., Fern{\'a}ndez Lorenzo M., et al., 2015, A\&A, 578, A110. doi:10.1051/0004-6361/201526016
\bibitem[\protect\citeauthoryear{Argudo-Fern{\'a}ndez et al.}{in prep.}]{Argudo2019} Argudo-Fern{\'a}ndez M., in prep.

\bibitem[\protect\citeauthoryear{Baldry, Glazebrook, \& Driver}{2008}]{Baldry2008} Baldry I.~K., Glazebrook K., Driver S.~P., 2008, MNRAS, 388, 945. doi:10.1111/j.1365-2966.2008.13348.x
\bibitem[\protect\citeauthoryear{Baldwin, Phillips, \& Terlevich}{1981}]{BPTdiagram} Baldwin J.~A., Phillips M.~M., Terlevich R., 1981, PASP, 93, 5. doi:10.1086/130766
\bibitem[\protect\citeauthoryear{Bao et al.}{2022}]{Bao2022} Bao M., Chen Y., Zhu P., Shi Y., Bizyaev D., Zhu L., Yang M., et al., 2022, ApJL, 926, L13. doi:10.3847/2041-8213/ac52ad
\bibitem[\protect\citeauthoryear{Belfiore et al.}{2017}]{Belfiore2017} Belfiore F., Maiolino R., Tremonti C., S{\'a}nchez S.~F., Bundy K., Bershady M., Westfall K., et al., 2017, MNRAS, 469, 151. doi:10.1093/mnras/stx789
\bibitem[\protect\citeauthoryear{Belfiore et al.}{2019}]{DAP2} Belfiore F., Westfall K.~B., Schaefer A., Cappellari M., Ji X., Bershady M.~A., Tremonti C., et al., 2019, AJ, 158, 160. doi:10.3847/1538-3881/ab3e4e
\bibitem[\protect\citeauthoryear{Belfiore et al.}{2016}]{Belfiore2016} Belfiore F., Maiolino R., Maraston C., Emsellem E., Bershady M.~A., Masters K.~L., Yan R., et al., 2016, MNRAS, 461, 3111. doi:10.1093/mnras/stw1234
\bibitem[\protect\citeauthoryear{Beom et al.}{in prep.}]{Beom2022} Beom, M., Walterbos, R.~A.~M., Bizyev, D., in prep.
\bibitem[\protect\citeauthoryear{Bertola, Galletta, \& Zeilinger}{1985}]{Bertola1985} Bertola F., Galletta G., Zeilinger W.~W., 1985, ApJL, 292, L51. doi:10.1086/184471
\bibitem[\protect\citeauthoryear{Bertola, Buson, \& Zeilinger}{1992}]{Bertola1992} Bertola F., Buson L.~M., Zeilinger W.~W., 1992, ApJL, 401, L79. doi:10.1086/186675
\bibitem[\protect\citeauthoryear{Bertola et al.}{1996}]{Bertola1996} Bertola F., Cinzano P., Corsini E.~M., Pizzella A., Persic M., Salucci P., 1996, ApJL, 458, L67. doi:10.1086/309924
\bibitem[\protect\citeauthoryear{Bertola \& Corsini}{1999}]{BertolaCorsini1999} Bertola F., Corsini E.~M., 1999, IAUS, 186, 149
\bibitem[\protect\citeauthoryear{Bettoni, Galletta, \& Oosterloo}{1991}]{Bettoni1991} Bettoni D., Galletta G., Oosterloo T., 1991, MNRAS, 248, 544. doi:10.1093/mnras/248.3.544
\bibitem[\protect\citeauthoryear{Bevacqua, Cappellari, \& Pellegrini}{2022}]{Becaxqua2022} Bevacqua D., Cappellari M., Pellegrini S., 2022, MNRAS, 511, 139. doi:10.1093/mnras/stab3732
\bibitem[\protect\citeauthoryear{Bizyaev et al.}{2014}]{Bizyaev14} Bizyaev D.~V., Kautsch S.~J., Mosenkov A.~V., Reshetnikov V.~P., Sotnikova N.~Y., Yablokova N.~V., Hillyer R.~W., 2014, ApJ, 787, 24. doi:10.1088/0004-637X/787/1/24
\bibitem[\protect\citeauthoryear{Bizyaev et al.}{2017}]{BWY17} Bizyaev D., Walterbos R.~A.~M., Yoachim P., Riffel R.~A., Fern{\'a}ndez-Trincado J.~G., Pan K., Diamond-Stanic A.~M., et al., 2017, ApJ, 839, 87. doi:10.3847/1538-4357/aa6979
\bibitem[\protect\citeauthoryear{Blanton et al.}{2011}]{NSA} Blanton M.~R., Kazin E., Muna D., Weaver B.~A., Price-Whelan A., 2011, AJ, 142, 31. doi:10.1088/0004-6256/142/1/31
\bibitem[\protect\citeauthoryear{Blanton et al.}{2017}]{Blanton2017} Blanton M.~R., Bershady M.~A., Abolfathi B., Albareti F.~D., Allende Prieto C., Almeida A., Alonso-Garc{\'\i}a J., et al., 2017, AJ, 154, 28. doi:10.3847/1538-3881/aa7567
\bibitem[\protect\citeauthoryear{Bottinelli et al.}{1985}]{Mrk273megamaser} Bottinelli L., Fraix-Burnet D., Gouguenheim L., Kazes I., Le Squeren A.~M., Patey I., Rickard L.~J., et al., 1985, A\&A, 151, L7
\bibitem[\protect\citeauthoryear{Braun, Walterbos, \& Kennicutt}{1992}]{BWK} Braun R., Walterbos R.~A.~M., Kennicutt R.~C., 1992, Natur, 360, 442. doi:10.1038/360442a0
\bibitem[\protect\citeauthoryear{Braun et al.}{1994}]{Braun1994} Braun R., Walterbos R.~A.~M., Kennicutt R.~C., Tacconi L.~J., 1994, ApJ, 420, 558. doi:10.1086/173586
\bibitem[\protect\citeauthoryear{Bryant et al.}{2019}]{Bryant2019} Bryant J.~J., Croom S.~M., van de Sande J., Scott N., Fogarty L.~M.~R., Bland-Hawthorn J., Bloom J.~V., et al., 2019, MNRAS, 483, 458. doi:10.1093/mnras/sty3122
\bibitem[\protect\citeauthoryear{Bundy et al.}{2015}]{manga_overview} Bundy K., Bershady M.~A., Law D.~R., Yan R., Drory N., MacDonald N., Wake D.~A., et al., 2015, ApJ, 798, 7. doi:10.1088/0004-637X/798/1/7
\bibitem[\protect\citeauthoryear{Bureau \& Chung}{2006}]{BureauChung} Bureau M., Chung A., 2006, MNRAS, 366, 182. doi:10.1111/j.1365-2966.2005.09840.x
\bibitem[\protect\citeauthoryear{Calzetti et al.}{2007}]{Calzetti2007} Calzetti D., Kennicutt R.~C., Engelbracht C.~W., Leitherer C., Draine B.~T., Kewley L., Moustakas J., et al., 2007, ApJ, 666, 870. doi:10.1086/520082
\bibitem[\protect\citeauthoryear{Cappellari et al.}{2007}]{Cappellari2007} Cappellari M., Emsellem E., Bacon R., Bureau M., Davies R.~L., de Zeeuw P.~T., Falc{\'o}n-Barroso J., et al., 2007, MNRAS, 379, 418. doi:10.1111/j.1365-2966.2007.11963.x
\bibitem[\protect\citeauthoryear{Cappellari}{2017}]{pPXF} Cappellari M., 2017, MNRAS, 466, 798. doi:10.1093/mnras/stw3020
\bibitem[\protect\citeauthoryear{Casanueva et al.}{2021}]{Casanueva2021} Casanueva C.~I., Lagos C. del P., Padilla N.~D., Davison T.~A., 2021, arXiv, arXiv:2110.00408
\bibitem[\protect\citeauthoryear{Cherinka et al.}{2019}]{PanSTARR} Cherinka B., Andrews B.~H., S{\'a}nchez-Gallego J., Brownstein J., Argudo-Fern{\'a}ndez M., Blanton M., Bundy K., et al., 2019, AJ, 158, 74. doi:10.3847/1538-3881/ab2634
\bibitem[\protect\citeauthoryear{Chen et al.}{2016}]{Chen2016} Chen Y.-M., Shi Y., Tremonti C.~A., Bershady M., Merrifield M., Emsellem E., Jin Y.-F., et al., 2016, NatCo, 7, 13269. doi:10.1038/ncomms13269
\bibitem[\protect\citeauthoryear{Chen et al.}{2019}]{Chen2019} Chen Y.-M., Shi Y., Wild V., Tremonti C., Rowlands K., Bizyaev D., Yan R., et al., 2019, MNRAS, 489, 5709. doi:10.1093/mnras/stz2494
\bibitem[\protect\citeauthoryear{Chung et al.}{2012}]{Chung2012} Chung A., Bureau M., van Gorkom J.~H., Koribalski B., 2012, MNRAS, 422, 1083. doi:10.1111/j.1365-2966.2012.20679.x
\bibitem[\protect\citeauthoryear{Ciri, Bettoni, \& Galletta}{1995}]{Ciri1995} Ciri R., Bettoni D., Galletta G., 1995, Natur, 375, 661. doi:10.1038/375661a0
\bibitem[\protect\citeauthoryear{Coccato et al.}{2013}]{Coccato2013} Coccato L., Morelli L., Pizzella A., Corsini E.~M., Buson L.~M., Dalla Bont{\`a} E., 2013, A\&A, 549, A3. doi:10.1051/0004-6361/201220460
\bibitem[\protect\citeauthoryear{Corsini}{2014}]{Corsini2014} Corsini E.~M., 2014, ASPC, 486, 51
\bibitem[\protect\citeauthoryear{Corsini \& Bertola}{1998}]{CorsiniBertola1998} Corsini E.~M., Bertola F., 1998, JKPS, 33, S574
\bibitem[\protect\citeauthoryear{Curti et al.}{2017}]{Curti2017} Curti M., Cresci G., Mannucci F., Marconi A., Maiolino R., Esposito S., 2017, MNRAS, 465, 1384. doi:10.1093/mnras/stw2766
\bibitem[\protect\citeauthoryear{Davies et al.}{2016}]{Davies2016} Davies L.~J.~M., Driver S.~P., Robotham A.~S.~G., Grootes M.~W., Popescu C.~C., Tuffs R.~J., Hopkins A., et al., 2016, MNRAS, 461, 458. doi:10.1093/mnras/stw1342
\bibitem[\protect\citeauthoryear{Davis et al.}{2011}]{Davis2011} Davis T.~A., Alatalo K., Sarzi M., Bureau M., Young L.~M., Blitz L., Serra P., et al., 2011, MNRAS, 417, 882. doi:10.1111/j.1365-2966.2011.19355.x
\bibitem[\protect\citeauthoryear{Dey et al.}{2019}]{DeCALS} Dey A., Schlegel D.~J., Lang D., Blum R., Burleigh K., Fan X., Findlay J.~R., et al., 2019, AJ, 157, 168. doi:10.3847/1538-3881/ab089d

\bibitem[\protect\citeauthoryear{Dom{\'\i}nguez S{\'a}nchez et al.}{2022}]{Dominguez_Sanchez2022} Dom{\'\i}nguez S{\'a}nchez H., Margalef B., Bernardi M., Huertas-Company M., 2022, MNRAS, 509, 4024. doi:10.1093/mnras/stab3089
\bibitem[\protect\citeauthoryear{Driver et al.}{2008}]{Driver2008} Driver S.~P., Popescu C.~C., Tuffs R.~J., Graham A.~W., Liske J., Baldry I., 2008, ApJL, 678, L101. doi:10.1086/588582
\bibitem[\protect\citeauthoryear{Drory et al.}{2015}]{Drory2015} Drory N., MacDonald N., Bershady M.~A., Bundy K., Gunn J., Law D.~R., Smith M., et al., 2015, AJ, 149, 77. doi:10.1088/0004-6256/149/2/77
\bibitem[\protect\citeauthoryear{Duckworth et al.}{2019}]{Duckworth2019} Duckworth C., Tojeiro R., Kraljic K., Sgr{\'o} M.~A., Wild V., Weijmans A.-M., Lacerna I., et al., 2019, MNRAS, 483, 172. doi:10.1093/mnras/sty3101
\bibitem[\protect\citeauthoryear{Duckworth, Tojeiro, \& Kraljic}{2020}]{Duckworth2020} Duckworth C., Tojeiro R., Kraljic K., 2020, MNRAS, 492, 1869. doi:10.1093/mnras/stz3575


\bibitem[\protect\citeauthoryear{Emsellem et al.}{2007}]{Emsellem2007} Emsellem E., Cappellari M., Krajnovi{\'c} D., van de Ven G., Bacon R., Bureau M., Davies R.~L., et al., 2007, MNRAS, 379, 401. doi:10.1111/j.1365-2966.2007.11752.x
\bibitem[\protect\citeauthoryear{Filippenko}{1985}]{Filippenko1985} Filippenko A.~V., 1985, ApJ, 289, 475. doi:10.1086/162909
\bibitem[\protect\citeauthoryear{Fisher, Illingworth, \& Franx}{1994}]{Fisher1994} Fisher D., Illingworth G., Franx M., 1994, AJ, 107, 160. doi:10.1086/116841

\bibitem[\protect\citeauthoryear{Galletta}{1996}]{Galletta1996} Galletta G., 1996, ASPC, 91, 429
\bibitem[\protect\citeauthoryear{Galletta}{1987}]{Galletta1987} Galletta G., 1987, ApJ, 318, 531. doi:10.1086/165389
\bibitem[\protect\citeauthoryear{Garc{\'\i}a-Burillo et al.}{2000}]{Garcia-Burillo2000} Garc{\'\i}a-Burillo S., Sempere M.~J., Combes F., Hunt L.~K., Neri R., 2000, A\&A, 363, 869
\bibitem[\protect\citeauthoryear{Garc{\'\i}a-Burillo, Sempere, \& Bettoni}{1998}]{Garcia-Burillo1998} Garc{\'\i}a-Burillo S., Sempere M.~J., Bettoni D., 1998, ApJ, 502, 235. doi:10.1086/305897
\bibitem[\protect\citeauthoryear{Goddy, Stark, \& Masters}{2020}]{Goddy2020} Goddy J., Stark D.~V., Masters K.~L., 2020, RNAAS, 4, 3. doi:10.3847/2515-5172/ab66bd
\bibitem[\protect\citeauthoryear{Gordon et al.}{2021}]{VALSSquick} Gordon Y.~A., Boyce M.~M., O'Dea C.~P., Rudnick L., Andernach H., Vantyghem A.~N., Baum S.~A., et al., 2021, ApJS, 255, 30. doi:10.3847/1538-4365/ac05c0
\bibitem[\protect\citeauthoryear{Graham \& Worley}{2008}]{GrahamWorley2008} Graham A.~W., Worley C.~C., 2008, MNRAS, 388, 1708. doi:10.1111/j.1365-2966.2008.13506.x
\bibitem[\protect\citeauthoryear{Gunn et al.}{2006}]{Gunn2006} Gunn J.~E., Siegmund W.~A., Mannery E.~J., Owen R.~E., Hull C.~L., Leger R.~F., Carey L.~N., et al., 2006, AJ, 131, 2332. doi:10.1086/500975


\bibitem[\protect\citeauthoryear{Kannappan \& Fabricant}{2001}]{KannappanFabricant2001} Kannappan S.~J., Fabricant D.~G., 2001, AJ, 121, 140. doi:10.1086/318027
\bibitem[\protect\citeauthoryear{Katkov et al.}{2011}]{Katkov} Katkov I., Chilingarian I., Sil'chenko O., Zasov A., Afanasiev V., 2011, BaltA, 20, 453. doi:10.1515/astro-2017-0318
\bibitem[\protect\citeauthoryear{Kautsch et al.}{2006}]{Kautsch2006} Kautsch S.~J., Grebel E.~K., Barazza F.~D., Gallagher J.~S., 2006, A\&A, 445, 765. doi:10.1051/0004-6361:20053981
\bibitem[\protect\citeauthoryear{Kennicutt et al.}{2009}]{Kennicutt2009} Kennicutt R.~C., Hao C.-N., Calzetti D., Moustakas J., Dale D.~A., Bendo G., Engelbracht C.~W., et al., 2009, ApJ, 703, 1672. doi:10.1088/0004-637X/703/2/1672
\bibitem[\protect\citeauthoryear{Kewley et al.}{2006}]{BPTlines} Kewley L.~J., Groves B., Kauffmann G., Heckman T., 2006, MNRAS, 372, 961. doi:10.1111/j.1365-2966.2006.10859.x

\bibitem[\protect\citeauthoryear{Khim et al.}{2020}]{Khim2020} Khim D.~J., Yi S.~K., Dubois Y., Bryant J.~J., Pichon C., Croom S.~M., Bland-Hawthorn J., et al., 2020, ApJ, 894, 106. doi:10.3847/1538-4357/ab88a9
\bibitem[\protect\citeauthoryear{Khim et al.}{2021}]{Khim2021} Khim D.~J., Yi S.~K., Pichon C., Dubois Y., Devriendt J., Choi H., Bryant J.~J., et al., 2021, ApJS, 254, 27. doi:10.3847/1538-4365/abf043
\bibitem[\protect\citeauthoryear{Kormendy \& Bender}{2012}]{KormendyBender2012} Kormendy J., Bender R., 2012, ApJS, 198, 2. doi:10.1088/0067-0049/198/1/2
\bibitem[\protect\citeauthoryear{Krajnovi{\'c} et al.}{2011}]{Krajnovic2011} Krajnovi{\'c} D., Emsellem E., Cappellari M., Alatalo K., Blitz L., Bois M., Bournaud F., et al., 2011, MNRAS, 414, 2923. doi:10.1111/j.1365-2966.2011.18560.x
\bibitem[\protect\citeauthoryear{Kuijken, Fisher, \& Merrifield}{1996}]{Kuijken1996} Kuijken K., Fisher D., Merrifield M.~R., 1996, MNRAS, 283, 543. doi:10.1093/mnras/283.2.543
\bibitem[\protect\citeauthoryear{Kumari et al.}{2019}]{Kumari2019} Kumari N., Maiolino R., Belfiore F., Curti M., 2019, MNRAS, 485, 367. doi:10.1093/mnras/stz366
\bibitem[\protect\citeauthoryear{Kumari et al.}{2021}]{Kumari2021} Kumari N., Maiolino R., Trussler J., Mannucci F., Cresci G., Curti M., Marconi A., et al., 2021, A\&A, 656, A140. doi:10.1051/0004-6361/202140757

\bibitem[\protect\citeauthoryear{Haynes et al.}{2000}]{Haynes2000} Haynes M.~P., Jore K.~P., Barrett E.~A., Broeils A.~H., Murray B.~M., 2000, AJ, 120, 703. doi:10.1086/301457
\bibitem[\protect\citeauthoryear{Haynes et al.}{2018}]{Haynes2018} Haynes M.~P., Giovanelli R., Kent B.~R., Adams E.~A.~K., Balonek T.~J., Craig D.~W., Fertig D., et al., 2018, ApJ, 861, 49. doi:10.3847/1538-4357/aac956
\bibitem[\protect\citeauthoryear{Hernquist \& Barnes}{1991}]{HernquistBernes1991} Hernquist L., Barnes J.~E., 1991, Natur, 354, 210. doi:10.1038/354210a0
\bibitem[\protect\citeauthoryear{Ho, Filippenko, \& Sargent}{1993}]{Ho1993} Ho L.~C., Filippenko A.~V., Sargent W.~L.~W., 1993, ApJ, 417, 63. doi:10.1086/173291
\bibitem[\protect\citeauthoryear{Holden et al.}{2012}]{Holden2012} Holden B.~P., van der Wel A., Rix H.-W., Franx M., 2012, ApJ, 749, 96. doi:10.1088/0004-637X/749/2/96
\bibitem[\protect\citeauthoryear{Masters et al.}{2019}]{Masters2019} Masters K.~L., Stark D.~V., Pace Z.~J., Phipps F., Rujopakarn W., Samanso N., Harrington E., et al., 2019, MNRAS, 488, 3396. doi:10.1093/mnras/stz1889
\bibitem[\protect\citeauthoryear{Murphy et al.}{2011}]{Murphy2011} Murphy E.~J., Condon J.~J., Schinnerer E., Kennicutt R.~C., Calzetti D., Armus L., Helou G., et al., 2011, ApJ, 737, 67. doi:10.1088/0004-637X/737/2/67

\bibitem[\protect\citeauthoryear{Jarrett et al.}{2011}]{Jarrett2011} Jarrett T.~H., Cohen M., Masci F., Wright E., Stern D., Benford D., Blain A., et al., 2011, ApJ, 735, 112. doi:10.1088/0004-637X/735/2/112
\bibitem[\protect\citeauthoryear{Jarrett et al.}{2013}]{Jarrett2013} Jarrett T.~H., Masci F., Tsai C.~W., Petty S., Cluver M.~E., Assef R.~J., Benford D., et al., 2013, AJ, 145, 6. doi:10.1088/0004-6256/145/1/6
\bibitem[\protect\citeauthoryear{Jarrett et al.}{2017}]{Jarrett2017} Jarrett T.~H., Cluver M.~E., Magoulas C., Bilicki M., Alpaslan M., Bland-Hawthorn J., Brough S., et al., 2017, ApJ, 836, 182. doi:10.3847/1538-4357/836/2/182
\bibitem[\protect\citeauthoryear{Jin et al.}{2016}]{Jin2016} Jin Y., Chen Y., Shi Y., Tremonti C.~A., Bershady M.~A., Merrifield M., Emsellem E., et al., 2016, MNRAS, 463, 913. doi:10.1093/mnras/stw2055

\bibitem[\protect\citeauthoryear{Law et al.}{2015}]{Law2015} Law D.~R., Yan R., Bershady M.~A., Bundy K., Cherinka B., Drory N., MacDonald N., et al., 2015, AJ, 150, 19. doi:10.1088/0004-6256/150/1/19
\bibitem[\protect\citeauthoryear{Law et al.}{2016}]{DRP} Law D.~R., Cherinka B., Yan R., Andrews B.~H., Bershady M.~A., Bizyaev D., Blanc G.~A., et al., 2016, AJ, 152, 83. doi:10.3847/0004-6256/152/4/83
\bibitem[\protect\citeauthoryear{Law et al.}{2021}]{DAP3} Law D.~R., Westfall K.~B., Bershady M.~A., Cappellari M., Yan R., Belfiore F., Bizyaev D., et al., 2021, AJ, 161, 52. doi:10.3847/1538-3881/abcaa2
\bibitem[\protect\citeauthoryear{Lacy et al.}{2020}]{VLASS} Lacy M., Baum S.~A., Chandler C.~J., Chatterjee S., Clarke T.~E., Deustua S., English J., et al., 2020, PASP, 132, 035001. doi:10.1088/1538-3873/ab63eb
\bibitem[\protect\citeauthoryear{Li et al.}{2021}]{Li2021} Li S.-. lin ., Shi Y., Bizyaev D., Duckworth C., Yan R.-. bin ., Chen Y.-. mei ., Bing L.-. ji ., et al., 2021, MNRAS, 501, 14. doi:10.1093/mnras/staa3618
\bibitem[\protect\citeauthoryear{Lovelace \& Chou}{1996}]{LovelaceChou} Lovelace R.~V.~E., Chou T., 1996, ApJL, 468, L25. doi:10.1086/310232
\bibitem[\protect\citeauthoryear{L{\'o}pez-Cruz et al.}{2019}]{LopezCruz2019} L{\'o}pez-Cruz O., Ibarra-Medel H.~J., S{\'a}nchez S.~F., Birkinshaw M., A{\~n}orve C., Barrera-Ballesteros J.~K., Falcon-Barroso J., et al., 2019, ApJL, 886, L2. doi:10.3847/2041-8213/ab5117

\bibitem[\protect\citeauthoryear{Peng et al.}{2010}]{Peng2010} Peng Y.-. jie ., Lilly S.~J., Kova{\v{c}} K., Bolzonella M., Pozzetti L., Renzini A., Zamorani G., et al., 2010, ApJ, 721, 193. doi:10.1088/0004-637X/721/1/193
\bibitem[\protect\citeauthoryear{Plana \& Boulesteix}{1996}]{PlanaBoulesteix} Plana H., Boulesteix J., 1996, A\&A, 307, 391
\bibitem[\protect\citeauthoryear{Pizzella et al.}{2004}]{Pizzella2004} Pizzella A., Corsini E.~M., Vega Beltr{\'a}n J.~C., Bertola F., 2004, A\&A, 424, 447. doi:10.1051/0004-6361:20047183

\bibitem[\protect\citeauthoryear{Rix et al.}{1992}]{Rix1992} Rix H.-W., Franx M., Fisher D., Illingworth G., 1992, ApJL, 400, L5. doi:10.1086/186635
\bibitem[\protect\citeauthoryear{Rix et al.}{1995}]{Rix1995} Rix H.-W.~R., Kennicutt R.~C., Braun R., Walterbos R.~A.~M., 1995, ApJ, 438, 155. doi:10.1086/175061
\bibitem[\protect\citeauthoryear{Rubin, Graham, \& Kenney}{1992}]{Rubin1992} Rubin V.~C., Graham J.~A., Kenney J.~D.~P., 1992, ApJL, 394, L9. doi:10.1086/186460
\bibitem[\protect\citeauthoryear{Rubin}{1994}]{Rubin1994} Rubin V.~C., 1994, AJ, 107, 173. doi:10.1086/116842

\bibitem[\protect\citeauthoryear{Sage \& Galletta}{1994}]{SageGalletta} Sage L.~J., Galletta G., 1994, AJ, 108, 1633. doi:10.1086/117184
\bibitem[\protect\citeauthoryear{Simard et al.}{2011}]{Simard2011} Simard L., Mendel J.~T., Patton D.~R., Ellison S.~L., McConnachie A.~W., 2011, ApJS, 196, 11. doi:10.1088/0067-0049/196/1/11
\bibitem[\protect\citeauthoryear{Sil'chenko, Moiseev, \& Shulga}{2010}]{Silchenko2010} Sil'chenko O.~K., Moiseev A.~V., Shulga A.~P., 2010, AJ, 140, 1462. doi:10.1088/0004-6256/140/5/1462
\bibitem[\protect\citeauthoryear{Smee et al.}{2013}]{BOSS} Smee S.~A., Gunn J.~E., Uomoto A., Roe N., Schlegel D., Rockosi C.~M., Carr M.~A., et al., 2013, AJ, 146, 32. doi:10.1088/0004-6256/146/2/32
\bibitem[\protect\citeauthoryear{Stark et al.}{2021}]{Stark2021} Stark D.~V., Masters K.~L., Avila-Reese V., Riffel R., Riffel R., Boardman N.~F., Zheng Z., et al., 2021, MNRAS, 503, 1345. doi:10.1093/mnras/stab566

\bibitem[\protect\citeauthoryear{Tadhunter et al.}{2018}]{Mrk273AGN} Tadhunter C., Rodr{\'\i}guez Zaur{\'\i}n J., Rose M., Spence R.~A.~W., Batcheldor D., Berg M.~A., Ramos Almeida C., et al., 2018, MNRAS, 478, 1558. doi:10.1093/mnras/sty1064
\bibitem[\protect\citeauthoryear{Tempel et al.}{2017}]{Tempel2017} Tempel E., Tuvikene T., Kipper R., Libeskind N.~I., 2017, A\&A, 602, A100. doi:10.1051/0004-6361/201730499
\bibitem[\protect\citeauthoryear{Thakar \& Ryden}{1996}]{ThakarRyden1996} Thakar A.~R., Ryden B.~S., 1996, ApJ, 461, 55. doi:10.1086/177037
\bibitem[\protect\citeauthoryear{Thakar et al.}{1997}]{Thakar1997} Thakar A.~R., Ryden B.~S., Jore K.~P., Broeils A.~H., 1997, ApJ, 479, 702. doi:10.1086/303915
\bibitem[\protect\citeauthoryear{Thakar \& Ryden}{1998}]{ThakarRyden1998} Thakar A.~R., Ryden B.~S., 1998, ApJ, 506, 93. doi:10.1086/306223

\bibitem[\protect\citeauthoryear{Walterbos, Braun, \& Kennicutt}{1994}]{Walterbos1994} Walterbos R.~A.~M., Braun R., Kennicutt R.~C., 1994, AJ, 107, 184. doi:10.1086/116843
\bibitem[\protect\citeauthoryear{Westfall et al.}{2019}]{DAP1} Westfall K.~B., Cappellari M., Bershady M.~A., Bundy K., Belfiore F., Ji X., Law D.~R., et al., 2019, AJ, 158, 231. doi:10.3847/1538-3881/ab44a2
\bibitem[\protect\citeauthoryear{Wiegert et al.}{2015}]{Wiegert2015} Wiegert T., Irwin J., Miskolczi A., Schmidt P., Mora S.~C., Damas-Segovia A., Stein Y., et al., 2015, AJ, 150, 81. doi:10.1088/0004-6256/150/3/81
\bibitem[\protect\citeauthoryear{Wright et al.}{2010}]{WISE} Wright E.~L., Eisenhardt P.~R.~M., Mainzer A.~K., Ressler M.~E., Cutri R.~M., Jarrett T., Kirkpatrick J.~D., et al., 2010, AJ, 140, 1868. doi:10.1088/0004-6256/140/6/1868


\bibitem[\protect\citeauthoryear{van de Voort et al.}{2015}]{vandeVoort2015} van de Voort F., Davis T.~A., Kere{\v{s}} D., Quataert E., Faucher-Gigu{\`e}re C.-A., Hopkins P.~F., 2015, MNRAS, 451, 3269. doi:10.1093/mnras/stv1217
\bibitem[\protect\citeauthoryear{van den Bergh}{2009}]{van_den_Bergh2009} van den Bergh S., 2009, ApJ, 702, 1502. doi:10.1088/0004-637X/702/2/1502
\bibitem[\protect\citeauthoryear{Vargas et al.}{2018}]{Vargas2018} Vargas C.~J., Mora-Partiarroyo S.~C., Schmidt P., Rand R.~J., Stein Y., Walterbos R.~A.~M., Wang Q.~D., et al., 2018, ApJ, 853, 128. doi:10.3847/1538-4357/aaa47f
\bibitem[\protect\citeauthoryear{Veilleux \& Osterbrock}{1987}]{VOdiagram} Veilleux S., Osterbrock D.~E., 1987, ApJS, 63, 295. doi:10.1086/191166
\bibitem[\protect\citeauthoryear{Wake et al.}{2017}]{Wake2017} Wake D.~A., Bundy K., Diamond-Stanic A.~M., Yan R., Blanton M.~R., Bershady M.~A., S{\'a}nchez-Gallego J.~R., et al., 2017, AJ, 154, 86. doi:10.3847/1538-3881/aa7ecc

\bibitem[\protect\citeauthoryear{Xu et al.}{2022}]{Xu2022} Xu H., Chen Y., Shi Y., Zhou Y., Bizyaev D., Bao M., Beom M., et al., 2022, MNRAS.tmp. doi:10.1093/mnras/stac354
\bibitem[\protect\citeauthoryear{Yan et al.}{2016a}]{Yan2016a} Yan R., Bundy K., Law D.~R., Bershady M.~A., Andrews B., Cherinka B., Diamond-Stanic A.~M., et al., 2016, AJ, 152, 197. doi:10.3847/0004-6256/152/6/197
\bibitem[\protect\citeauthoryear{Yan et al.}{2016b}]{Yan2016b} Yan R., Tremonti C., Bershady M.~A., Law D.~R., Schlegel D.~J., Bundy K., Drory N., et al., 2016, AJ, 151, 8. doi:10.3847/0004-6256/151/1/8
\bibitem[\protect\citeauthoryear{Zeng, Wang, \& Gao}{2021}]{Zeng2021} Zeng G., Wang L., Gao L., 2021, MNRAS, 507, 3301. doi:10.1093/mnras/stab2294

\end{thebibliography}


\bsp	
\label{lastpage}
\end{document}